\documentclass[pra,onecolumn,superscriptaddress,showpacs,nofootinbib]{revtex4-1}
\usepackage{graphicx}

\usepackage{bm}
\usepackage{amssymb}
\usepackage{amsmath}
\usepackage{amstext}
\usepackage{latexsym}

\newcommand{\bra}[1]{\langle #1|}
\newcommand{\ket}[1]{|#1\rangle}
\newcommand{\ex}[1]{\langle #1 \rangle}

\begin{document}


\title{Enhancement  of laser cooling by the use of magnetic gradients}


\author{A. Albrecht}
\email[]{andreas.albrecht@uni-ulm.de}

\author{A. Retzker}
\affiliation{
Institut f\"ur Theoretische Physik, Universit\"at Ulm, 89069 Ulm, Germany
}
\author{C. Wunderlich}
\affiliation{
Fachbereich Physik, Universit\"at Siegen, D-57068 Siegen, Germany
}
\author{M. B. Plenio}
\affiliation{
Institut f\"ur Theoretische Physik, Universit\"at Ulm, 89069 Ulm, Germany
}

\date{\today}

\begin{abstract}
We present a laser cooling scheme for trapped ions and atoms using a combination of laser couplings and a magnetic gradient field. In a Schrieffer-Wolff transformed picture, this setup cancels the carrier and blue sideband terms completely (up to first order in the Lamb-Dicke parameter), resulting in an improved cooling behaviour compared to  standard cooling schemes in the Lamb-Dicke regime (e.g. sideband cooling) and allowing cooling to the  vibrational ground state. A condition for optimal cooling rates is presented and the cooling behaviour for different Lamb-Dicke parameters and spontaneous decay rates is discussed. Cooling rates of  one order of magnitude less than the trapping frequency are achieved using the new cooling method. Furthermore the scheme exhibits fast rates and low final populations even for significant deviations from the optimal parameters and moreover provides good cooling rates also in the multi-particle case.
\end{abstract}

\pacs{}

\maketitle


\section{Introduction}
Cooling of atoms/ions is an important precondition for quantum logic and computation \cite{Wineland2005,sasura01}, wherein an initial ground state preparation of the vibrational state is required, for high resolution spectroscopy beyond the Doppler limitations and for the analysis of low temperature phenomena (e.g. BEC condensates). Cooling by light (laser cooling) for atoms and ions was originally proposed in \cite{haensch74,wineland75}.  For trapped ions, sideband cooling \cite{stenholm86,wineland78}  is the standard cooling method, achieved by tuning the laser frequency to the red sideband of the bare atomic frequency, which leads to a resonant cooling contribution whereas heating  is only induced by off-resonant carrier and blue-sideband excitations. To obtain well-resolved sidebands requires the linewidth of the atomic transition to be much smaller than the trapping frequency $\Gamma\ll\nu$ (strong confinement). Furthermore, taking into account that off-resonant heating excitations increase with increasing laser intensity   the coupling Rabi-frequency has to be smaller than the trapping frequency $\Omega\ll\nu$ (weak coupling) \cite{sasura01}. Cooling to lower temperatures was reached using the concept of Raman sideband cooling \cite{lindberg86}, applicable for atoms \cite{kerman00} and ions \cite{lee96,monroe95}, which allows the realisation of very small cooling linewidths by Raman two-photon transitions. Further improvements on the way to reach the ground state are achieved with dark state concepts based on eliminating off-resonant excitations via quantum interference.  In EIT cooling \cite{morigi00,roos00} the concept of electromagnetically induced transparency \cite{fleischhauer05} is used to eliminate the carrier transition, which gives the main heating contribution in the sideband schemes. It uses a three level system coupled by two lasers, wherein one of the lasers (the stronger one) creates the absorption spectrum,  a Fano like profile with a zero at the carrier transition and a peak at the red sideband transition. Therefore stronger couplings can be realised, no strong confinement is required due to the three level nature of the system and a large cooling bandwidth (i.e. frequency range for cooling) allows simultaneous cooling of several modes. An alternative approach, based on the Stark shift gate \cite{jonathan00}, is the Stark shift cooling \cite{retzker07}, cancelling the carrier transition for a correctly tuned Rabi frequency, which is of the order of the trapping frequency. This is achieved in the dressed state picture, i.e. taking the Stark shift of the laser interaction (originating from the zeroth order in the Lamb-Dicke parameter contributions of the resonant coupling laser) into account. The robust laser cooling scheme \cite{cerrillo10} combines the methods of EIT and Stark shift cooling resulting in a dark state for the combination of EIT and Stark shift Hamiltonian contribution and leading to an elimination of both the carrier transition and the blue sideband heating transition by interference.

The scheme presented here also eliminates the carrier and blue sideband transitions in lowest order in the Lamb-Dicke parameter, using an atomic three level system, a combinations of four laser beams and  applying a magnetic field gradient. The application of an inhomogeneous magnetic field to trapped ions resulting in a position dependent resonance shift  and mediating the coupling between internal and external degrees of freedom (e.g. allowing the coupling by microwaves with a non-vanishing effective Lamb-Dicke parameter) was discussed in \cite{mintert01,wunderlich03}.  For our method only the proper choice of the gradient strength and the phase relations of the laser beams   form the conditions for a pure red sideband interaction up to first order in the Lamb-Dicke parameter. \par
The paper is organized as follows: In section\,\ref{one} the cooling scheme is presented and the influence of the magnetic field gradient is analysed. It turns out, that in a Schrieffer-Wolff transformed picture, both the carrier as well as the blue sideband transition are eliminated by interference and thus cooling to the vibrational ground state in zeroth order is achieved. In section\,\ref{three} the resonance condition for the cooling process is derived. Moreover it is shown that the maximal cooling amplitude is limited by the trapping frequency, a fact that arises from the amplitudes of the dressed states, which are tuned to resonance. Section\,\ref{four} presents an analytic formula for the cooling rate obtained by the method of adiabatic elimination and valid in the weak coupling case. Indeed a comparison to numerical results shows a good agreement in the range of small Rabi frequencies. In section\,\ref{five} the dependence of the cooling rate on the spontaneous decay rate as well as on the magnitude of the Lamb-Dicke parameter is analysed. Furthermore the dependence of the cooling rate and final population on phase fluctuations of the coupling lasers is presented resulting in the conclusion that the scheme, albeit becoming slower, still provides fast cooling and low final populations even for relatively large deviations from the optimal phase relation. Section\,\ref{six} deals with the cooling of multiple vibrational modes, i.e. with multiple trapped ions. Only for the center of mass mode a complete elimination of heating terms (carrier and blue sideband) is possible; however the strength of the heating terms for other modes is smaller than in standard cooling schemes.

\section{Cooling scheme and Hamiltonian}\label{one}
\begin{figure*}
\includegraphics[scale=0.6]{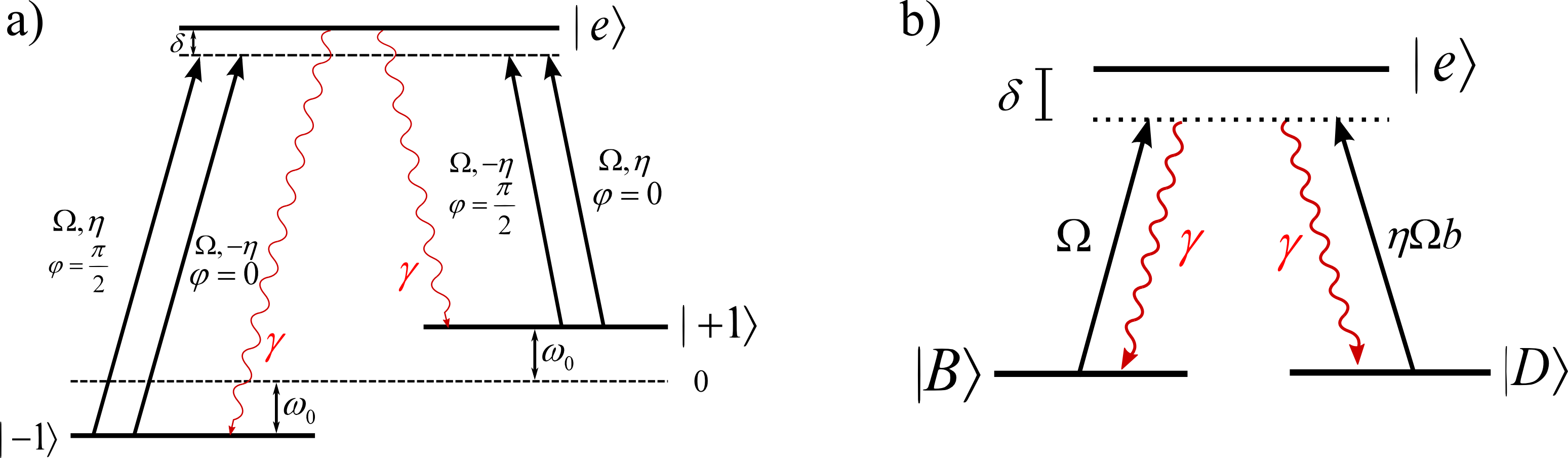}
\caption{\label{fsetup} Cooling scheme: a) Coupling scheme in the standard basis. Each of the Zeeman split ground state levels is coupled to the excited level $\ket{e}$ by two lasers with Rabi frequency $\Omega$, Lamb Dicke parameter $\eta$, phase $\varphi$ and detuning $\delta$.  Note that the external magnetic field gives, beside the position independent $\omega_0$ splitting, an additional position dependent energy shift via the magnetic gradient contribution. b) Scheme in dressed state basis after Schrieffer-Wolff transformation. See also eq.\,(\ref{ham4}).  }
\end{figure*}

The scheme presented here is based on a complete elimination of the blue sideband heating terms in a Schrieffer-Wolff transformed picture (see eq.\,(\ref{sw1}) and (\ref{sw2}) below) in lowest order in the Lamb-Dicke parameter, which is achieved by a magnetic gradient field together with a four laser coupling scheme. It requires an atomic three level system consisting of a (fast decaying) excited state $\ket{e}$ and a Zeeman split ground state $\ket{+1}$ and $\ket{-1}$ induced by the external position dependent magnetic field. Each of the Zeeman split levels is coupled by two lasers with Rabi frequency $\Omega$  and opposite Lamb-Dicke parameters $\eta$ ($\eta=k\,\sqrt{\hbar/(2\,m\,\nu)}\,\cos\theta$, with $\theta$ the angle of the laser beam to the trap axis and $k$ the absolute value of the wavevector), obtained by irradiating the beams from opposite directions, to the excited level $|e\rangle$ having a phase difference $\varphi=\pi/2$ (see fig.\ref{fsetup}).\\
The ion is assumed to be trapped in an harmonic trapping potential with frequency $\nu$ and for simplicity only the motion along the trap axis is taken into account. Moreover, the expansion up to first order in terms of the Lamb-Dicke parameter requires the atom/ion to be localized on the scale provided by the laser wavelength (Lamb-Dicke limit, $\eta \ll 1$). 
The position dependent magnetic field $B(x)$ has the form
\begin{equation} B(x) \approx B_0+\left.\dfrac{\mathrm{d}B}{\mathrm{d}x}\right|_{x=0}\,x    \end{equation}   
and higher terms are not present or can be neglected in the following expansion in terms of the effective Lamb-Dicke parameter respectively.
The corresponding energy shift due to the magnetic Zeeman splitting is (with the position operator $x=\sqrt{\frac{\hbar}{2\,m\,\nu}}(b+b^\dagger)$, wherein $b^\dagger$ and $b$ are the phonon creator and annihilator operators respectively) given by
\begin{equation}\label{en_split} \Delta E=-\mu\,B=\hbar\,\omega_0\,m_s+\eta_{eff}\,\hbar\,\nu\,\left(b+b^\dagger\right)\,m_s   \end{equation}
with $\mu$ the magnetic moment, $m_s$  the magnetic quantum number and $\eta_{eff}=\frac{\sqrt{\hbar}}{\sqrt{2\,m\,\nu}}\frac{\omega_0}{\nu B_0}\,\frac{\partial B}{\partial x}$, providing a coupling between the spatial position and the internal energy, i.e. phonons and internal atomic states. The effective Lamb Dicke parameter has a physical meaning as the energy difference between the left and right bound of the ground state wavefunction extent in units of $\hbar\,\nu$. The energy shift of the homogeneous part is given by $\hbar\,\omega_0=-g\,\mu_B\,B_0$, wherein $g$ denotes the g-factor and $\mu_B$ the Bohr magneton.
Thus the Hamiltonian of the system (cp. to fig.\,\ref{fsetup}) is given by (setting $\hbar=1$)
\begin{equation}\label{ham1}\begin{split}    
H=&\nu\,b^\dagger\,b+\omega_e\,\ket{e}\bra{e}+\omega_0\,\left(\ket{+1}\bra{+1}-\ket{-1}\bra{-1}  \right)
		+\eta_{eff}\,\nu\,\left(\ket{+1}\bra{+1} -\ket{-1}\bra{-1} \right)\,\left(b+b^\dagger  \right)\\
		&+\Omega\,\left(\ket{e}\bra{+1}+\ket{+1}\bra{e}  \right)\cdot\left[\cos\left( k\,x-\omega_L\,t \right)+\cos\left(-k\,x-\omega_L\,t+\pi/2  \right)\right]\\
		&+\Omega\,\left(\ket{e}\bra{-1}+\ket{-1}\bra{e}  \right)\cdot\left[\cos\left(-k\,x-\omega^*_L\,t  \right)+\cos\left(k\,x-\omega^*_L\,t+\pi/2  \right)  \right]
\end{split}
\end{equation}
with $\omega_e$ being the energy of the upper level $\ket{e}$.
In order to end up with a time independent Hamiltonian we assume for the detuning: $\delta=\omega_e-\omega_0-\omega_L=\omega_e+\omega_0-\omega_L^*$, wherein $\omega_L$ is the frequency of the first laser pair (coupling $\ket{+1}$ to $\ket{e}$) and $\omega_L^*$ the frequency of the second pair respectively (coupling $\ket{-1}$ to $\ket{e}$).
In an interaction picture, expanding the Hamiltonian up to the first order in the Lamb-Dicke parameter $\eta$ and applying the rotating wave approximation, the Hamiltonian takes the form
\begin{equation}
\begin{split}
\label{ham2}
H=&\nu\,b^\dagger b+\delta\,\ket{e}\bra{e}+\eta_{eff}\,\nu \left(\ket{+1}\bra{+1}-\ket{-1}\bra{-1}\right)\,\left(b+b^\dagger \right)\\
  &+\left[\dfrac{\Omega}{2}\,(1+i)\left(\ket{e}\bra{+1}+\ket{e}\bra{-1}\right)\right.
  \left .+\eta\,\dfrac{\Omega}{2}\,(1+i)\,\left(b+b^\dagger  \right)\left(\ket{e}\bra{+1}-\ket{e}\bra{-1}\right)+\text{h.c.}\right].
\end{split}
\end{equation}
In order to cancel  the blue sideband heating contributions completely, it is important to adjust the magnetic gradient field in a way so that
\begin{equation}\label{ass1} \eta=\eta_{eff}     \end{equation}
as will be explained later. Values for the necessary magnetic gradient to fulfill condition (\ref{ass1}) for Ytterbium-172 and Calcium-40 ions are given in table\,\ref{tabelle1}. \\
\begin{table}
\caption{\label{tabelle1} Magnetic gradients needed to fulfill condition\,(\ref{ass1}) for different trapping frequencies. Values are given for the $S_{1/2}$ to $P_{1/2}$ transitions of Ytterbium-172 ions (369\,nm) and Calcium-40 ions (397\,nm). In this case the magnetic gradient is given by $\frac{\partial B}{\partial x}=\frac{2\,\nu\,\hbar}{(-g_e)\,\mu_B}\,k\cos\theta$, wherein $\theta$ denotes the angle between the laser beams and the trap axis ($\eta=\sqrt{\hbar\,k^2/(2\,m\,\nu)}\,\cos\theta$), which in the following is assumed to be zero and $g_e$ denotes the electron g-factor.}
\renewcommand{\arraystretch}{1.5}
\begin{tabular}{|l||c|c||c|c|} \hline
 &\multicolumn{2}{|c||}{$\boldsymbol{^{172}Yb^+}$}         &\multicolumn{2}{c|}{$\boldsymbol{^{40}Ca^+}$}\\\hline\hline
\textbf{Trapping frequency} $\boldsymbol{[\nu/(2\,\pi)]/Hz}$ & $\qquad 5\cdot10^5 \qquad $& $\qquad 1\cdot10^6 \qquad$ & $\qquad 5\cdot 10^5 \qquad$ & $\qquad 1\cdot 10^6 \qquad$\\\hline
\textbf{Magnetic gradient} $\boldsymbol{\frac{\partial B}{\partial x}}$\textbf{/(T/m)} & 607.6 & 1215 & 564.7 & 1130 \\\hline
\textbf{Lamb-Dicke parameter} $\boldsymbol{\eta}$  & 0.13 & 0.09 & 0.25 & 0.18  \\\hline
\end{tabular}
\end{table}
The basic idea of the cancellation mechanism is that $\sigma_x (b+b^+) + i \sigma_y (b-b^+)$ generates the red sideband and cancels the blue one.
Since the phonon coupling introduces a term that is proportional to $b+b^+$ and the magnetic gradient introduces a term that is proportional   to $i(b-b^+),$ we can expect that if the phases are adjusted appropriately  we can get a complete interference of the blue sideband.

The effect of the magnetic field is best seen by applying a unitary Schrieffer-Wolff transformation $U$\,\cite{schrieffer66} to the Hamiltonian (\ref{ham2})
\begin{equation}\label{sw1}  U=\mathrm{exp}\left\{-\eta\,\left(\ket{+1}\bra{+1}-\ket{-1}\bra{-1}\right)\,\left(b-b^\dagger\right)\right\}      \end{equation}
resulting in the following transformations
\begin{equation} \label{sw2}
\begin{split}
b \qquad &\rightarrow \quad b-\eta\,\left( \ket{+1}\bra{+1}-\ket{-1}\bra{-1}  \right)\\
\ket{+1}  \qquad &\rightarrow \quad \ket{+1}\,\mathrm{exp}\left[-\eta\left(b-b^\dagger\right)\right]    \\
\ket{-1} \qquad &\rightarrow \quad \ket{-1}\,\mathrm{exp}\left[\phantom{-}\eta\left(b-b^\dagger\right)\right]  \,\, .
\end{split}
\end{equation} 

Application of the Schrieffer-Wolff transformation to the Hamiltonian (\ref{ham2}) and keeping only terms up to first order in $\eta$ we get (fig.\,\ref{fsetup}b, note that $\eta$ is equal to $\eta_{eff}$):
\begin{equation}\label{ham4}
H'=U\,H\,U^{\dagger}
  =\nu\,b^\dagger\,b+\delta\,\ket{e}\bra{e}
   +\left[\dfrac{\Omega}{\sqrt{2}}\,(1+i)\,\ket{e}\bra{B}\right.
   +\left.\eta\,\Omega\,\sqrt{2}\,(1+i)\,b\,\ket{e}\bra{D}+\text{h.c.}\right].
\end{equation}
Herein we have defined:
\begin{equation}
\begin{split}
\ket{B}&=\dfrac{1}{\sqrt{2}}\,\left(\ket{+1}+\ket{-1}\right),\\
\ket{D}&=\dfrac{1}{\sqrt{2}}\,\left(\ket{+1}-\ket{-1}\right).
\end{split}
\end{equation}

Note that the first order expansion terms of the Schrieffer-Wolff transformation ($\eta_{eff}$) of the zeroth order term in $\eta$ (Lamb-Dicke-Parameter) Hamiltonian contributions (see \ref{ham2}) lead to a cancellation of the first order in $\eta$ blue sideband laser interaction contributions under the assumption (\ref{ass1}).

The complete time evolution in the Schrieffer-Wolff transformed picture is given by
\begin{equation}\label{p_evol1} \dfrac{\partial}{\partial t}\rho=-i\left[ H',\rho \right]+\mathcal{L}^d \,\rho.  \end{equation}
The Liouvillian Superoperator $\mathcal{L}^d$ describes the spontaneous decay from the excited level $\ket{e}$ to the levels $\ket{+1}$ and $\ket{-1}$. It exhibits the following form:
\begin{equation} \mathcal{L}^d=\mathcal{L}_0^d+\mathcal{L}_1^d \,.    \end{equation}
$\mathcal{L}_0^d\,\rho$ describes the general spontaneous decay terms in zeroth order in the Lamb-Dicke Parameter:
\begin{equation}\label{spdec1} \mathcal{L}_0^d\,\rho=\sum_{i=\{(+1),(-1)\}}\dfrac{\gamma_i}{2}\,\Bigl[2\, \ket{i}\bra{e}\rho \ket{e}\bra{i}-\ket{e}\bra{e}\,\rho-\rho\,\ket{e}\bra{e}   \Bigr]\,.    \end{equation}
$\mathcal{L}_1^d\,\rho$ is the Schrieffer-Wolff transformation contribution of the spontaneous decay term (in first order in $\eta$):
\begin{equation}\label{spdec2} \mathcal{L}_1^d\,\rho = \sum_{i=\{(+1),(-1)\}} \gamma_i\,(-1)^i\,\eta\,\Bigl[\left(b-b^\dagger\right) \bm{,} \ket{i}\bra{e}\rho\ket{e}\bra{i} \Bigr]\,. \end{equation}
Note that recoil only contributes in order $\eta^4$, leading to a very small correction to the cooling rate (decreasing it) due to the dark state nature of the cooling scheme and therefore can be neglected for small $\eta$ (which was tested by numerical simulations).  This is due to the fact, that the probability for being in the excited state is given by $\eta^2$ and the spontaneous decay including recoil  from this state is a process in second order in $\eta$ and therefore recoil only contributes with a strength $\propto \eta^4 $. 

If the final phonon population $\ket{n=0}$ is reached, the system will be decoupled from the laser interaction and will end up in the state $\ket{D,n=0}$ (`` dark state''). Note that due to the presence of only red sideband terms the phonon state $\ket{n=0}$ will always be reached (unless $\Omega=0$).  In the original (i.e. non Schrieffer-Wolff transformed) picture, this final state has the form:
\begin{equation}\label{state_origianl} U^\dagger\,\ket{D,n=0}=\ket{D,n=0}-\eta\,\ket{B,n=1}+\mathcal{O}\left(\eta^2\right)\,.    \end{equation}
Therefore the presented scheme cools to the vibrational ground state (n=0) in zeroth order in $\eta$. Furthermore the Schrieffer-Wolff transformation preserves pure states and thus does not increase the entropy (which remains zero).  A simple (projective) measurement of the internal state $\ket{D}$ as well as a unitary rotation could further improve the achieved final population, that is they would result in a state with $n=0$ up to second order in $\eta$.\par

\section{Conditions for optimal cooling rates and limitations}\label{three}
\begin{figure}
\includegraphics[scale=0.16]{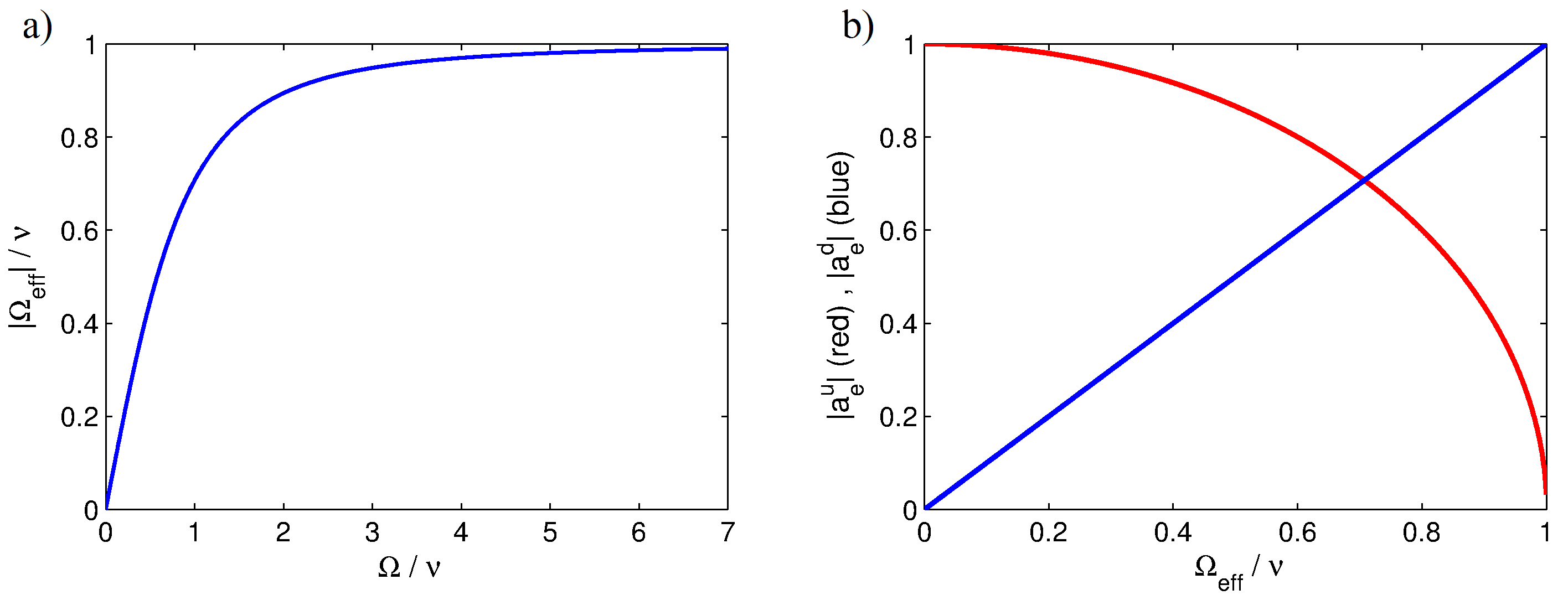}
\caption{\label{foeff} a) $|\Omega_{eff}|/\nu$ is plotted over the Rabi frequency $\Omega$ in units of the trapping frequency $\nu$. It turns out that the maximal achievable $|\Omega_{eff}|$  is limited to the trapping frequency $\nu$ (see also eq.\,(\ref{oeff1})).
b) Absolute value of $\ket{e}$ amplitudes in the dressed states; the red curve represents the resonant state. The minimal amplitude determines the dynamics for a given $\Omega_{eff}$. }
\end{figure}

At first sight it might seem possible to increase the cooling rate by an arbitrary amount just by increasing the laser Rabi frequency $\Omega$. However a detailed analysis of the $\Omega\,\ket{e}\bra{B}$-coupling terms (cp. Hamiltonian (\ref{ham4})) shows that this is not the case. These terms are changing the $\ket{e}$-state  (shifting it and mixing it with the $\ket{B}$-state) and therefore crucially influence the cooling term  ($\sim \eta\Omega b \ket{e}\bra{D}$) . Optimal rates are achieved for tuning the cooling interaction in resonance with one of the dressed states of the $\Omega\,\ket{e}\bra{B}$ interaction. The Hamiltonian (\ref{ham4}) in the dressed state basis takes the form
\begin{equation}
\begin{split}
H'=&\nu\,b^\dagger\,b+\omega_u\,\ket{u}\bra{u}+\omega_d\,\ket{d}\bra{d}+\eta\,\Omega\,\sqrt{2}\,(1+i)\,b\,\left\{\left(a^u_e\ket{u}+a^d_e\,\ket{d}\right)\bra{D}+\text{h.c.}      \right\}
\end{split}   
\end{equation}
wherein $\ket{u}$ and $\ket{d}$ is the dressed state basis (each having contributions of $\ket{e}$ and $\ket{B}$ states), $\omega_u$ and $\omega_d$ are the energies of those states (i.e. eigenvalues of the diagonalization procedure) 
\begin{equation}\label{ds1}  \omega_{u/d}=\dfrac{1}{2}\,\left(\delta\pm\sqrt{\delta^2+4\,\Omega^2}\right)   \end{equation}
and $a_e^u$ and $a_e^d$ the amplitudes of $\ket{u}$ and $\ket{d}$ in $\ket{e}$ respectively (see also eq.\,(\ref{eamp1})), i.e. the excited level $\ket{e}$ expressed in the dressed state basis is given by
\begin{equation}\label{eamp1} \ket{e}=a^u_e\,\ket{u}+a^d_e\,\ket{d}\,.    \end{equation}
In the interaction picture with respect to the first three Hamiltonian contributions, the resonance condition (for tuning the $\ket{u}$ state to resonance) is easily seen to be fulfilled by setting $\omega_u=\nu $.
 This procedure leads to the following resonance condition for optimal cooling rates:
\begin{equation}\label{resonance_cond1}  \delta=\dfrac{\nu^2-\Omega^2}{\nu}\,.    \end{equation}
Consequently the detuning is - for a given trapping and laser Rabi frequency - uniquely determined in the optimal cooling case. \\
To see how the magnitude of the cooling term increases with increasing $\Omega$ it is useful to define an ``effective Rabi frequency`` $\Omega_{eff}$.
 If $\ket{u}$ corresponds to the resonant state (the one that has been tuned to resonance), then $\Omega_{eff}$ is defined as
\begin{equation}\label{oeffdef} \Omega_{eff}=\Omega\,a^u_e     \end{equation}
  and is a direct measure of the  strength of the resonant (and therefore relevant) cooling contribution. In case of fulfilling condition (\ref{resonance_cond1}) it follows that
\begin{equation}\label{oeff1}  |\Omega_{eff}|=\dfrac{\nu\,|\Omega|}{\sqrt{\nu^2+\Omega^2}}    \end{equation}
(see figure \ref{foeff}a).
From equation (\ref{oeff1}) it is clearly seen, that it is impossible to increase the cooling term contribution above a limiting upper value, or more precisely, $\Omega_{eff}$ is limited to a maximal value, which corresponds to the trapping frequency $\nu$ ($|\Omega_{eff}|_{max}=\nu$). Therefore the trapping frequency is responsible for an upper limit to the maximal achievable cooling rate.

A second limitation arises from the decay rate $\gamma$ and the contributions of the excited state $\ket{e}$ in the new dressed state basis. Note that spontaneous decay to both dressed states and the $\ket{D}$ state takes place. The cooling process relies on the decay to the $\ket{D}$ state, and thus - to keep the process going - it is necessary that the other states decay fast enough, i.e. have a sufficient large $\ket{e}$ contribution. Otherwise the system will be trapped (at least for a certain amount of time) in one of the dressed states and thus the cooling rate is reduces significantly. For equal decay rates $\gamma$ in (\ref{spdec1}),(\ref{spdec2}), equal $\ket{e}$ contributions in the dressed states would lead to the optimal case when taking only the spontaneous decay influence into account. The  $\ket{e}$-amplitudes in the dressed states for the resonance case (\ref{resonance_cond1}) are plotted in figure \ref{foeff}b. Note that the minimal amplitude determines the dynamics. After a point of equal $\ket{e}$-amplitudes in the dressed states, the minimal $\ket{e}$-amplitude decreases again for further increasing $\Omega_{eff}$, which sets the optimal rate to a value $\Omega_{eff}<\nu$, depending on the relation between $\gamma$ and $\nu$.

\section{Rate equation formula and comparison to the numerical results}\label{sec_formula}\label{four}
\begin{figure}
\includegraphics[scale=0.9]{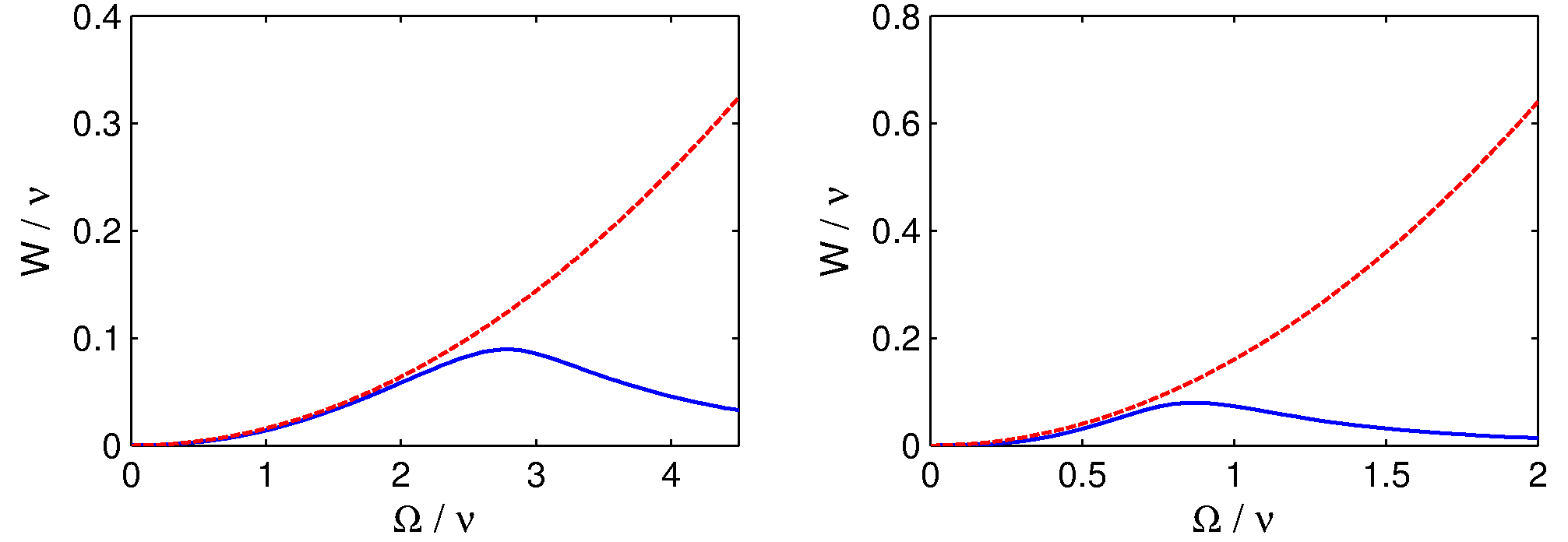}
\caption{\label{frcomp} Cooling rate $W$ as a function of the Rabi frequency $\Omega$ both in units of the trapping frequency $\nu$ for $\eta=0.1$ and $\gamma=5\,\nu$ (left plot) / $\gamma=0.5\,\nu$ (right plot). The detuning $\delta$ for a given $\Omega$ is chosen to fulfill the resonance condition eq.\,(\ref{resonance_cond1}). The blue line is the rate obtained by numerical simulations, the red dashed line the one calculated using the rate equation formula (\ref{rate2}).}
\end{figure}

The procedure described in \cite{cirac92,lindberg86} allows to calculate an analytic formula for the cooling rate by adiabatically eliminating the internal degrees of freedom under the assumption that the cooling dynamics is much slower than the internal dynamics. Moreover it is assumed that the trapping frequency is large compared to the cooling dynamics (i.e. $\eta\,\Omega \ll\nu$), which allows a projection on the phonon $\ket{n}\bra{n}$ subspace; i.e., a derivation of a rate equation. This second condition is only  fulfilled for small values of $\Omega$ in our case (otherwise the cooling term is of the order of the trapping frequency or for further increasing $\Omega$ even bigger). 
The rate $W$ can be written in the form $W=A_--A_+$ and the final steady state population is given by $\langle n \rangle_{ss}=A_+/(A_--A_+)$, wherein $A_+$ /$ A_-$- are the transition rates for increasing or lowering the phonon number respectively. Those quantities consist of two time correlation functions of internal operators, which can be calculated using the quantum regression theorem and the optical Bloch equations.  It follows that, due to the fact that there exist no heating terms in first order in the Lamb-Dicke Parameter, $A_+$ is equal to zero and thus $W=A_-$ and the final phonon population is given by $\langle n  \rangle_{ss}=0$.\\
Following the procedure given in \cite{cirac92}, we get for the rate $W$ (see appendix for the calculation)
\begin{equation}\label{rate_formula}  W= \dfrac{8\,\eta^2\,\Omega^2\,\gamma\,\nu^2}{\gamma^2\,\nu^2+\left[(\delta_L-\nu)\,\nu+\Omega^2) \right]^2}  \,  ,  \end{equation} 
where the rate $W$ is defined via:
\begin{equation}\label{rate_def}  \dfrac{\mathrm{d}}{\mathrm{d}t} \langle n\rangle =-W\, \langle n\rangle\, .   \end{equation}
This rate (\ref{rate_formula}) gets optimal if the term in brackets is equal to zero and the resulting condition is equivalent to the resonance condition given in eq.\,(\ref{resonance_cond1}). In that case, the rate takes the following form:
\begin{equation}\label{rate2}  W=\dfrac{8\,\eta^2\,\Omega^2}{\gamma} \,.   \end{equation}  
The fact that the rate is proportional to $1/\gamma$ is a direct consequence of the adiabatic elimination in high $\gamma$. More precisely, the rate $W$ is proportional  to the spontaneous decay rate $\gamma$ of the excited level times the probability $p_u$ for achieving this level. The latter quantity is proportional to $1/\gamma^2$, i.e. the probability for reaching the excited dressed state (here: $\ket{u}$) decreases with increasing dissipation.  Therefore the rate is given by: $W\propto (1/\gamma^2)\,\gamma=1/\gamma$.
However it does not mean that the optimal achievable rate is getting slower for increasing $\gamma$, because it is possible to reach higher final values for $\Omega$ due to the fact, that the (internal) dynamics is less limited by the decreasing $\ket{e}$-amplitude of the resonant state, which occurs for increasing $\Omega$ beyond a point of equal dressed state amplitudes. 
Figure\,\ref{frcomp} shows the rate obtained by numerical simulations and that one calculated using eq.\,(\ref{rate2}) for different Rabi frequencies. As expected, the rate formula describes the rate quite well in the limit of small $\Omega$ values whereas it fails completely in the opposite limit (when the cooling contribution is getting stronger and the conditions used for the formula derivation are no longer fulfilled).

\section{Simulation and results}\label{five}
\begin{figure}
\includegraphics[scale=0.2]{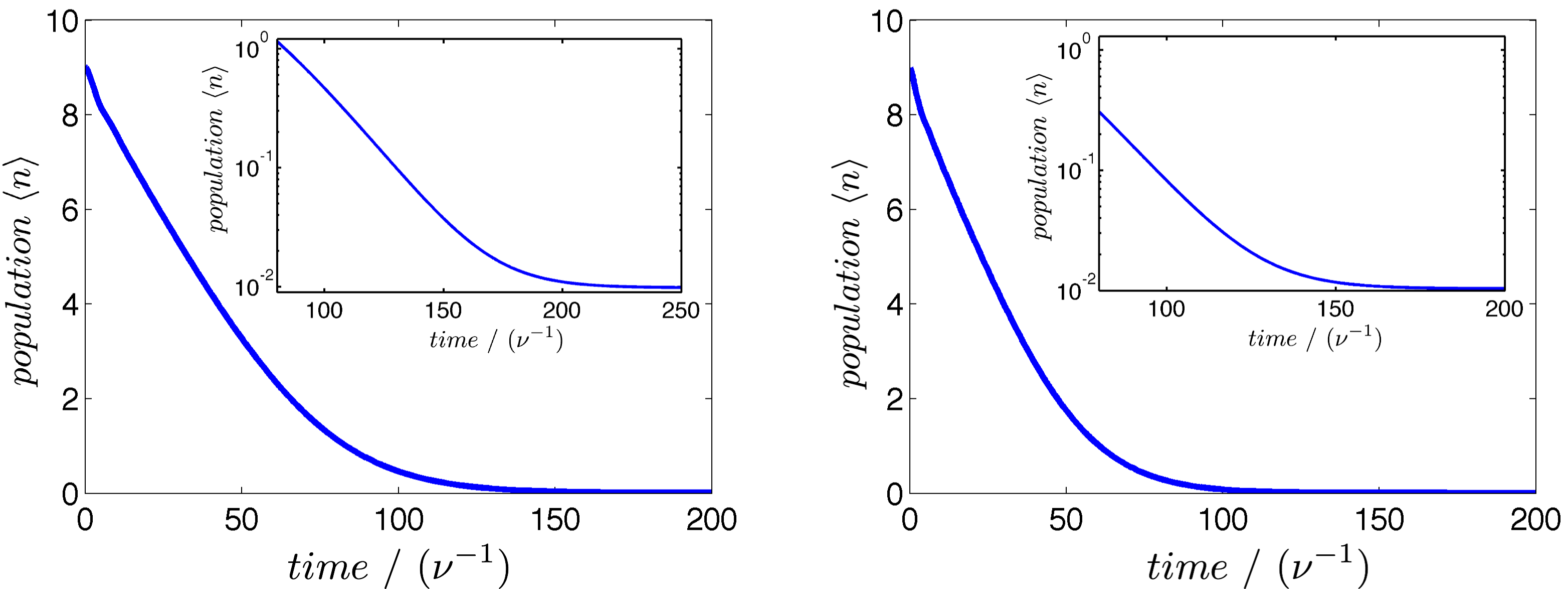}
\caption{\label{fsim} Average population $\langle n \rangle$ vs. time in units of the trapping frequency $\nu$ for optimally chosen cooling parameters. The left plot describes the optimal case for $\gamma=0.5\,\nu$ ($\Omega=0.85\,\nu, \delta=0.28\,\nu$, $\eta=0.1$), the right plot for  $\gamma=5\,\nu$ ($\Omega=2.8\,\nu$, $\delta=-6.84\,\nu$, $\eta=0.1$). The insets show the final cooling stage of the corresponding figures using a logarithmic scale for the population, which clearly reveals the final population of the order of $\eta^2$. }
\end{figure}
Figure \ref{fsim} shows optimal cooling cycles for different spontaneous decay rates ($\gamma=0.5\,\nu<\nu$ and $\gamma=5\,\nu>\nu$). In the first case ($\gamma<\nu$, the rate is of the magnitude of $\gamma$) the optimal cooling rate is achieved for $\Omega_{eff}\approx\nu$ (cp. to eq.\,(\ref{resonance_cond1})) , i.e. the case of equal $\ket{e}$-amplitudes in the dressed states. That is what one would expect, because with further increasing $\Omega_{eff}$ to its maximal value $\nu$ there is one dressed state with a small $\ket{e}$ amplitude and therefore the slower spontaneous decay from this state leads to a decreasing rate despite the increasing strength of the cooling term contribution. In the $\gamma=5\,\nu, \gamma>\nu$, $W\ll\gamma$ (cooling rate) case, the spontaneous decay rate is much faster than the cooling dynamics. Therefore, as expected, it is possible to increase $\Omega_{eff}$ beyond the ''equal amplitude`` situation and thus a slightly higher effective Rabi frequency can be realized. The spontaneous decay time of the dressed states is much less limiting in this case. Note that this higher $\Omega$ is needed to compensate the influence of the quantum Zeno effect, which reduces the rate compared to the $\gamma=0.5\,\nu$ case for a given $\Omega$.
The optimal achieved rates $W$ (see also eq.\,(\ref{rate_def})) are $0.08\,\nu$ ($\Omega=0.85\,\nu, \Omega_{eff}=0.65\,\nu$) for the $\gamma=0.5\,\nu$ case and $0.09\,\nu$ ($\Omega=2.8\,\nu, \Omega_{eff}=0.94\,\nu$) for the $\gamma=5\,\nu$ case respectively.
Rates for different values of $\Omega$ are presented in figure \ref{frcomp}.

\begin{figure}
\includegraphics[scale=0.6]{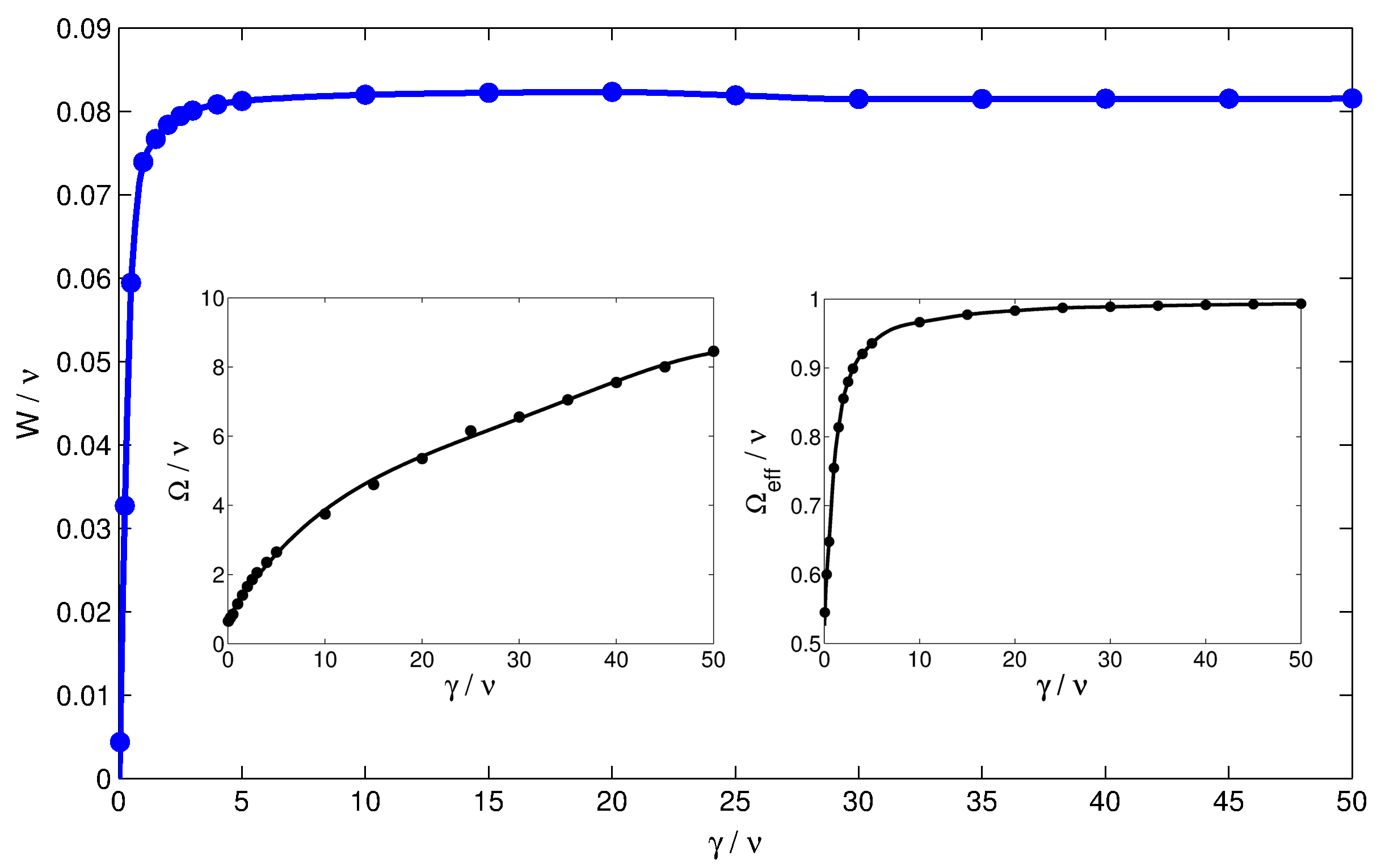}
\caption{\label{fdiffgamma} Cooling rate $W$ in units of the trapping frequency $\nu$ for different values of the spontaneous decay rate $\gamma$ (in units of the trapping frequency $\nu$). The inset shows the optimal value of $\Omega$ (left) and $\Omega_{eff}$ (right) for a given $\gamma$. While the rate increases fast in regions where the cooling is limited by the spontaneous decay rate (small values of $\gamma$) it saturates for high spontaneous decay rates, where the rate is limited by the limitation of $\Omega_{eff}$ (see also fig.\,\ref{foeff}). Values have been calculated for $\eta=0.1$, optimal values for the Rabi frequency and the detuning chosen according to the resonance condition eq.\,(\ref{resonance_cond1}).}
\end{figure}

The dependence of the cooling rate $W$ on the spontaneous decay rate $\gamma$ is plotted in fig.\,\ref{fdiffgamma}. As discussed before, for higher spontaneous decay rates higher values of $\Omega$ are possible because of less restrictions due to the spontaneous decay and therefore the rate increases with increasing $\gamma$. For small values of $\gamma$, when the cooling dynamics is limited by the spontaneous decay rate and not by the $\Omega_{eff}$ coupling, the rate increases significantly with increasing $\gamma$. For high values of $\gamma$ the rate converges to a limiting value for the rate. In this case the cooling dynamics is clearly limited by the fact, that $\Omega_{eff}$ reaches its maximal value $\nu$ (the trapping frequency (\ref{oeff1}), see right inset of fig.\,\ref{fdiffgamma} for the corresponding $\Omega_{eff}$ values). Thus further increasing $\Omega$ in the region $\Omega_{eff} \lesssim\nu$ for increasing $\gamma$ (which is indeed necessary to compensate for the higher quantum Zeno effect) leads to a very small increasing of $\Omega_{eff}$ and therefore leads to a saturation of the rate with $\Omega_{eff}$ converging to its maximal value.

\begin{figure}
\includegraphics[scale=0.35]{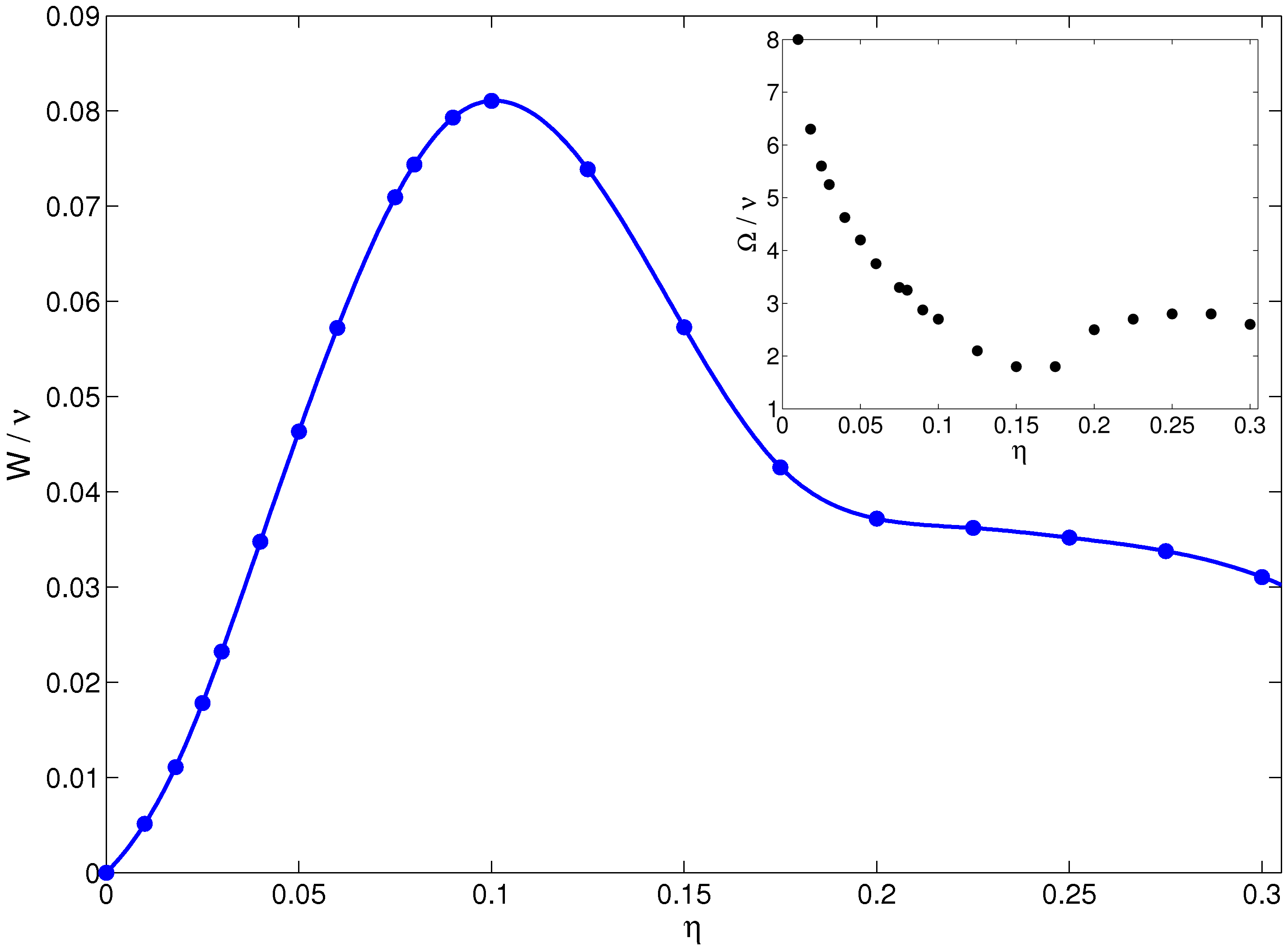}
\caption{\label{fdiffeta} Cooling rate $W$ in units of the trapping frequency $\nu$ for different values of the Lamb-Dicke parameter $\eta$. The corresponding $\Omega$ dependence (for optimal rates) is shown in the inset. For small values of $\eta$ (when the influence of higher order heating terms is small) the rate scales approximately linear in $\eta$. All values have been calculated for a fixed spontaneous decay rate $\gamma=5\,\nu$ and optimal detunings according to the resonance condition\,(\ref{resonance_cond1}), including terms up to second order in $\eta$.}
\end{figure}

The dependence on $\eta$ (for otherwise fixed parameters) is shown in fig.\,\ref{fdiffeta}. For $\eta\ll 1$, i.e. in the regime where heating terms of higher orders in $\eta$ can be neglected, the rate scales approximately linearly with  increasing Lamb-Dicke parameter $\eta$. Note that the rate equation (\ref{rate2}) includes a quadratic contribution of $\eta$, but also the optimal Rabi frequency depends on $\Omega$, scaling approximately $\propto 1/\sqrt{\eta}$ and thus leading to an approximately linear scaling\footnote{More exactly, numerical simulations show that the approximate scaling is given by $W\propto \eta^{1.28}$ and $\Omega\propto\eta^{-0.45}$.} of the rate dependence on $\eta$. Thus decreasing $\eta$ leads to increasing  $\Omega$ for the optimal condition, which can be explained as compensating for the quantum Zeno effect although it deteriorates the spontaneous decay amplitude behaviour. By increasing $\eta$ to higher values it reaches some optimal point and decreases again afterwards due to the fact that the heating terms in second order (and by further increasing also in higher orders) in $\eta$ become more and more important.

The cooling scheme concept is designed for the Lamb-Dicke regime ($\eta\ll1$), because only in this regime higher order contributions can be neglected. Numerical simulations for higher Lamb-Dicke parameters show that the (minimal) final population always increases with increasing Lamb-Dicke parameter. While the final population for $\eta=0.1$ is given by $\langle n\rangle = 0.01$ corresponding to $\eta^2$ as expected according to\,(\ref{state_origianl}), it is given by $\ex{n}\simeq0.06$ for $\eta=0.2$, $\ex{n}\simeq0.64$ for $\eta=0.3$ and $\ex{n}\simeq4.5$ for $\eta=0.6$ ($\Omega=13\,\nu$, $\gamma=5\,\nu$). Therefore only in the Lamb-Dicke regime the goal to achieve the vibrational ground state can be achieved. With respect to the new final population the cooling exhibits for higher Lamb-Dicke parameters in average again an exponential decay, leading e.g. for the above described $\eta=0.6$ case to a cooling rate of $W\simeq 0.006\,\nu$. Note however that the cooling rate and the final temperature are not optimized simultaneously for the same parameters, which especially for high Lamb-Dicke parameters leads to significant differences in the final populations and rates. The fact that both the final population and rate are worse for increasing Lamb-Dicke parameters underlines the restriction to the Lamb-Dicke regime.\vspace{1ex}

\begin{figure}
\includegraphics[scale=0.6]{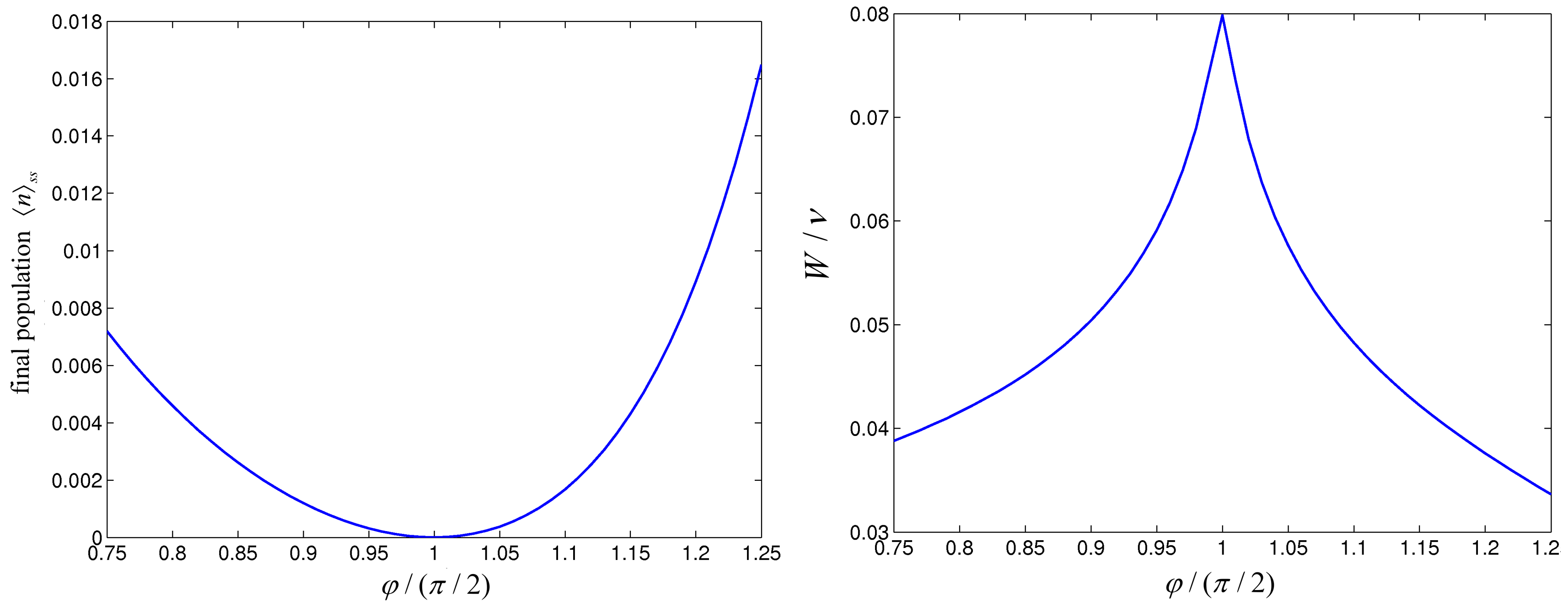}
\caption{\label{phase}
The left plot shows the final population $\langle n \rangle_{ss}$, the right plot the rate $W$ in units of $\nu$ for different phase differences $\varphi$ between the two coupling laser pairs. Note that $\varphi=\pi/2$ corresponds to the optimal case allowing the complete elimination of heating terms in first order in $\eta$. The simulation is performed in the Schrieffer-Wolff transformed picture and an Hamiltonian expansion up to second order in the Lamb-Dicke parameter with the parameters  $\Omega=2.8\,\nu, \delta=-6.84\,\nu$ (according to the resonance condition eq.\,(\ref{resonance_cond1})), $\eta=0.1$ and $\gamma=5\,\nu$ is used. Note that the final population corresponds to the population in the transformed picture. 
}
\end{figure}

{Robustness to phase fluctuations --- } The presented scheme relies on destructive interference and thus may be fragile to dephasing, i.e., fluctuations in the phase difference between the laser pairs around the optimal $\pi/2$ value. The numerical analysis of the phase fluctuation is shown in fig. \ref{phase}.  It can be seen that even at fluctuations which are larger than $20\%$ the population is still less than $10^{-2}$ and the rates decrease only by a factor of $2.$

The proposed cooling scheme is faster  than previously proposed continuous cooling schemes. The fastest continuous scheme that was implemented is the EIT cooling scheme \cite{roos00} which cooled at a rate of $10^{-4} \nu$. The robust cooling scheme \cite{cerrillo10} can in principle reach a rate of $10^{-2} \nu.$ In contrary the proposed cooling scheme can reach rates which are one order of magnitude less than the trap frequency.

\section{Multimode cooling} \label{six}
The presented cooling scheme was tested for the case of several trapped ions. Furthermore the cooling scheme, i.e. the laser coupling of internal and external degrees of freedom as presented in section\,\ref{one}, is applied to one of the trapped ions (named ion\,1 in the following). It is advantageous to express the Hamiltonian in terms of normal modes, wherein the creator and annihilator of the n-th normal mode is denoted by $a_n^\dagger$ and $a_n$ respectively. In contrast, the ones of the i-th local mode are denoted by $b_i$ and $b_i^\dagger$. Expanding equation\,(\ref{ham2}) to the multi particle case leads to the Hamiltonian
\begin{equation}\label{sevm1}\begin{split} H=&\delta\,\ket{e}\bra{e}+\sum_{n} \nu_n\,a_n^\dagger\,a_n+ \eta\,\nu\,\left[\ket{+1}\bra{+1}-\ket{-1}\bra{-1} \right]\,\left(b_1+b_1^\dagger  \right) \\
					  &+\left[\dfrac{\Omega}{\sqrt{2}}\,(1+i)\,\ket{e}\bra{B}
					  +\eta\,\dfrac{\Omega}{\sqrt{2}}\,(1+i)\,\ket{e}\bra{D}\left(b_1+b_1^\dagger \right)+\text{h.c.}\right]+\mathcal{O}(\eta^2)\, .
 \end{split} \end{equation}
Herein $\nu_n$ describes the frequencies of the corresponding normal modes, the summation over $n$ equals a summation over all normal modes and all internal states correspond to those of ion\,1. With the transformation $x_i=\sum_{n}M_i^n\,Q_n$ between the local coordinates $\{x_i\}$ and the normal coordinates $\{Q_n\}$ the local operators in (\ref{sevm1}) can be expressed in terms of normal ones using the relation ($x_i=\sqrt{\hbar/(2m\nu)}\,(b_i+b_i^\dagger)$)
\begin{equation}\label{sevm3}  \left(b_i+b_i^\dagger \right)=\sum_{n}\,M_i^n\,\sqrt{\frac{\nu}{\nu_n}}\,\left(a_n+a_n^\dagger \right)     \end{equation}
and thus the Hamiltonian\,(\ref{sevm1}) reads
\begin{equation}\label{sevm4}\begin{split} H=&\delta\,\ket{e}\bra{e}+\sum_{n}\left[\nu_n\,a_n^\dagger\,a_n +\eta_n'\,\nu\,\left[\ket{+1}\bra{+1}-\ket{-1}\bra{-1}\right]\, \left(a_n+a_n^\dagger  \right) \right]\\ &+\left[\dfrac{\Omega}{\sqrt{2}}\,(1+i)\,\ket{e}\bra{B}
+\sum_{n}\eta_n' \,\dfrac{\Omega}{\sqrt{2}}\,(1+i)\,\ket{e}\bra{D}\,(a_n+a_n^\dagger)+\text{h.c.}\right]+\mathcal{O}(\eta^2)\, .
 \end{split}\end{equation}
wherein $\eta_n'=\eta\,M_1^n\,\sqrt{\nu/\nu_n}$.\\
Again, the effect of the magnetic gradient terms is best seen by moving to a Schrieffer-Wolff transformed picture in which the magnetic gradient displacements of the individual normal mode oscillators are removed. This transformation is given by 
\begin{equation}\label{sevm8}\begin{split}  U'&=\text{exp}\left[-\sum_{n}\tilde{\eta}_n\,[\ket{+1}\bra{+1}-\ket{-1}\bra{-1}]\,(a_n-a_n^\dagger)\right] \\
		&=\prod_n\,\text{exp}\left[-\tilde{\eta}_n\,[\ket{+1}\bra{+1}-\ket{-1}\bra{-1}]\,\left(a_n-a_n^\dagger \right)\right]	 \end{split}\end{equation}
with $\tilde{\eta}_{n}=\eta_n'\,\nu/\nu_n=\eta\,M_1^n\,(\nu/\nu_n)^{3/2}$. Note that this transformation differs from the transformation used in sect.\,\ref{one} what can easily be checked  by inserting (\ref{sevm3}) into (\ref{sw1}).
Applying this transformation (cp. to eq.\,(\ref{sw2})) to the Hamiltonian\,(\ref{sevm1}) leads to $H'=U'\,H\,U'^\dagger$
\begin{equation}\label{sevm10}\begin{split} H'
  =&\delta \,\ket{e}\bra{e}+\sum_{n}\nu_n\,a_n^\dagger\,a_n+\biggl[\dfrac{\Omega}{\sqrt{2}}\,(1+i)\,\ket{e}\bra{B}\\  
		      &+\eta\,\dfrac{\Omega}{\sqrt{2}}\,(1+i)\,\ket{e}\bra{D}\,\sum_{n}\,M_1^n\,\sqrt{\dfrac{\nu}{\nu_n}}\left(\left[1+\dfrac{\nu}{\nu_n}  \right]\,a_n+\left[1-\dfrac{\nu}{\nu_n} \right]\,a_n^\dagger  \right)+\text{h.c.}\biggr]+\mathcal{O}(\eta^2)\,.
 \end{split}\end{equation}
Consequently, the amplitudes of the blue and red sideband terms for a specific mode $n$ depend on the relation of the trapping frequency $\nu$ to the normal mode frequency $\nu_n$. A complete cancellation of blue sideband heating terms is only possible for the center of mass mode (COM), because only in this case $\nu_{COM}=\nu$. For higher modes heating terms are always present, although they are off-resonant by tuning the laser frequency to the red sideband. In general, for increasing $\nu_n$ the heating amplitude increases and the cooling amplitude decreases, which leads to equal amplitudes for $\nu_n\to\infty$. The latter case corresponds to the situation in cooling schemes that do not cancel blue sideband terms, e.g. EIT cooling and sideband cooling. Therefore the presented cooling scheme results in an improved cooling behaviour with respect to those schemes also in the multiparticle case albeit not being able to cool other modes than the COM-mode without heating terms. \par
\begin{figure}
\includegraphics[scale=0.2]{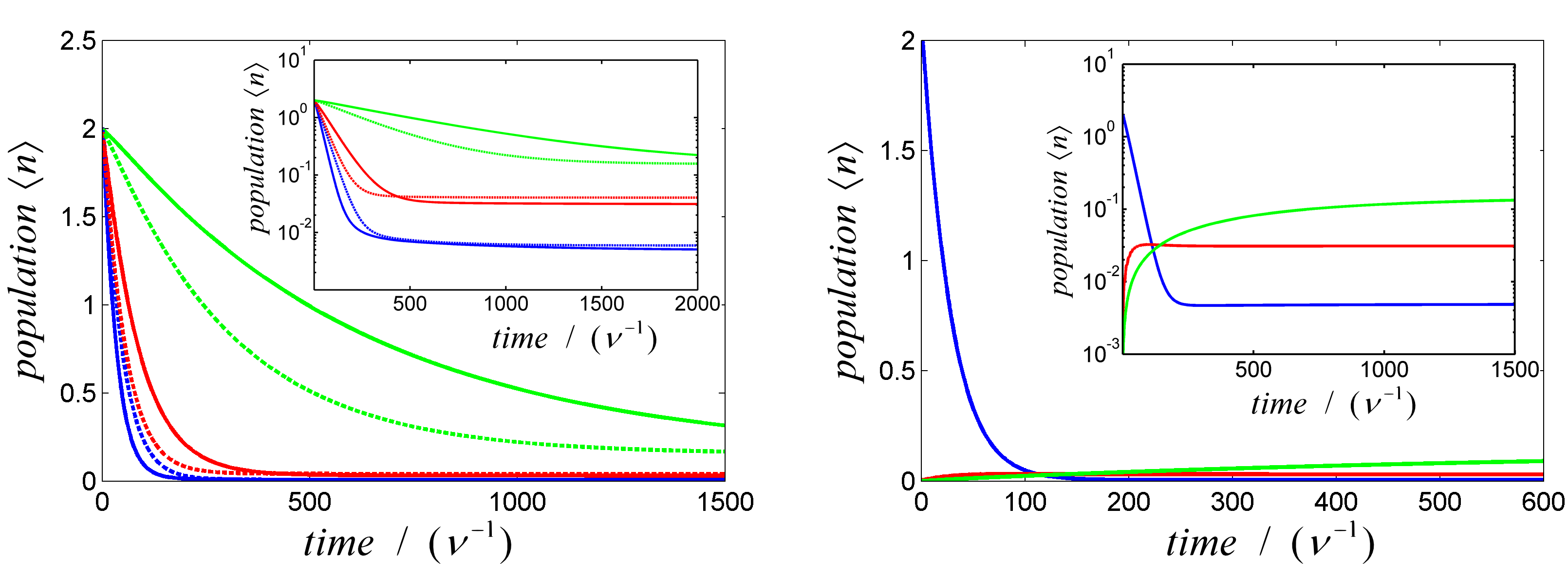}
\caption{\label{bsevmod3} Cooling of  three trapped ions. The line colours blue, red and green correspond to the single modes with increasing frequency in this order. The simulation was performed in the original (non transformed) picture using a second order in $\eta$ Hamiltonian expansion and the values $\Omega=2.8\,\nu, \gamma=5\,\nu$ and $\eta=0.1$. The insets show the corresponding cooling processes with a logarithmic scaling of the population. The left plot shows the cooling of the three modes all starting with an initial population $\langle n \rangle=2$. The detuning $\delta$ was chosen in a way to set the red sideband terms of the first and second mode to resonance respectively, corresponding to continuous and dashed lines. Continuous lines describe the situation when the laser is tuned to resonance with the red sideband of the center of mass mode and dashed lines the analogous situation with the breathing mode tuned to resonance (see resonance condition eq.\,(\ref{resonance_cond1})).  The right plot shows the heating of the excited modes during the cooling process of the center of mass mode, which is set to resonance. The final populations are: $\langle n_1 \rangle=0.005$, $\langle n_2 \rangle=0.031$ and $\langle n_3 \rangle=0.14$. }
\end{figure}
The cooling of the individual modes (COM mode (blue), first excited mode (breathing mode, red) and second excited mode (egyptian mode, green)) for the case of three ions is presented in fig.\,\ref{bsevmod3}. For three ions \cite{sasura01} $\vec{M}^1=(1/\sqrt{3},-1/\sqrt{2},1/\sqrt{6})$, $\nu_1=\nu$, $\nu_2=\sqrt{3}\,\nu$, $\nu_3=\sqrt{29/5}\,\nu$. The left plot shows the cooling starting with an initial population $\langle n \rangle=2$ for each mode.  Due to the fact that in the higher excited modes heating terms are always present while in the COM mode those contributions are eliminated completely in first order in $\eta$, the COM mode is always cooled with a faster rate also in the case when the cooling terms of any of the excited modes is set to resonance, i.e. the COM cooling transitions are off-resonant. The right plot shows the heating of the non-resonant excited modes originally in $\langle n \rangle =0$ during the resonant cooling process of the center of mass mode starting from $\langle n_{1} \rangle=2$. The modes are cooled/heated to final populations of $\langle n_1 \rangle=0.005$, $\langle n_2 \rangle=0.031$ and $\langle n_3 \rangle=0.14$. That is the heating turns out to be very small and increases with $\nu_n$ as expected from eq.\,(\ref{sevm10}).

\section{Conclusion and comments}
We presented a cooling scheme in the Lamb-Dicke regime, which completely eliminates the influence of the blue sideband heating transitions as well as the carrier transitions in lowest order in the Lamb-Dicke parameter,  resulting in a good cooling behaviour and providing cooling rates of the order of one magnitude below the trapping frequency. Albeit the Lamb-Dicke regime required for the Hamiltonian expansion and the resonance condition limit the cooling rate well below the trapping frequency, the achieved rates are faster than in comparable existing schemes working in the same regime (e.g. EIT cooling\,\cite{morigi00}, Robust cooling scheme\,\cite{cerrillo10}).  This is achieved by a proper choice of the magnetic gradient strength whose influence was analysed in a Schrieffer-Wolff transformed picture.  Optimal cooling conditions were derived together with a formula valid in the weak laser coupling case. The dependence on the Lamb-Dicke  parameter as well as on the spontaneous decay rate was presented. Moreover the cooling behaviour under deviations around the optimal phase differences between the coupling laser pairs was analysed showing that deviations of about 20 percent only reduce the rates by a factor of two and the final population is still of the order of $\eta^2$, both very satisfying values compared to other cooling schemes working in the Lamb-Dicke regime. Furthermore an analysis of the multiparticle case revealed good cooling behaviour also for several modes albeit not being able to eliminate the carrier and blue sideband terms completely up to first order in $\eta$ in other modes than the center of mass mode.  \par

\begin{acknowledgments}
Support by the European Commission (STREPs HIP and PICC), the Alexander von Humboldt foundation, the Deutsche Forschungsgemeinschaft, and secunet AG is acknowledged.
\vspace{5ex}
\end{acknowledgments}
\appendix
\section{Rate equation derivation}
In this appendix, the calculation leading to the rate equation formula (\ref{rate_formula}) is presented.
The master equation (\ref{p_evol1}) up to first order in $\eta$ can be written in the following form
\begin{equation}\label{app1}  \dfrac{\mathrm{d}\rho}{\mathrm{d}t}=\mathcal{L}_0\,\rho+\left(\mathcal{L}_1^1+\mathcal{L}_1^2 \right)\,\rho  \end{equation}
wherein the lower index indicates the order in the Lamb-Dicke parameter. The Liouvillians are given by
\begin{equation}\label{app2} \mathcal{L}_0\,\rho =\left(\mathcal{L}_{0,ext}+\mathcal{L}_{0,int} \right) \,\rho=-i\,\left[H_{0,ext},\rho \right]+\left[-i\left[ H_{0,int},\rho \right]+\mathcal{L}_0^d\,\rho   \right] \end{equation}
\begin{equation}\label{app3} \mathcal{L}_1^1\,\rho=-i\,\left[H_1,\rho  \right]  , \quad \mathcal{L}_1^2\,\rho=\mathcal{L}_1^d\,\rho  \end{equation}
with the definitions of the spontaneous decay contributions $\mathcal{L}_0^d\,\rho$ and $\mathcal{L}_1^d\,\rho$ given in eq.\,(\ref{spdec1}) and eq.\,(\ref{spdec2}) respectively. The complete first order Liouvillian $\mathcal{L}_1$ is defined as the sum of the two first order contributions, i.e. $\mathcal{L}_1=\mathcal{L}_1^1+\mathcal{L}_1^2$. The zero and first order Hamiltonian contributions (cp. to eq.\,(\ref{ham4})) are
\begin{equation}\label{app4}\begin{split}  H_{0,ext}&=\nu\,b^\dagger b \\  H_{0,int}&=\delta_1\,\ket{e}\bra{e}+\left\{\frac{\Omega}{\sqrt{2}}\,(1+i)\,\ket{e}\bra{B}+\text{h.c.}\right\}   \end{split}   \end{equation}
\begin{equation}\label{app5} H_1=b\,\eta\,\Omega\,\sqrt{2}\,(1+i)\,\ket{e}\bra{D}+\text{h.c.} =\sum_{n\in\{ 1,2\}}R_n\,F_n  \end{equation}
where $R_n$ are pure external operators and $F_{n}$ pure internal ones: $R_{1}:=b$, $R_{2}:=R_{1}^\dagger$, $F_{1}:=\eta\,\Omega\,\sqrt{2}\,(1+i)\,\ket{e}\bra{D}$ and $F_{2}=F_{1}^\dagger$.\\
The projector $\mathcal{P}$ on the relevant subspace is given by
\begin{equation}\label{app6}\begin{split} \mathcal{P}\rho&=\mathcal{P}^e\,\mathcal{P}^i \rho=\mu\,\rho_{ss}\\
					    \mathcal{P}^i\,\rho&=\underset{t\to\infty}{\lim} e^{\mathcal{L}_{0,int}\,t}\,\rho=\rho_{ss}\,\text{tr}_I(\rho)\\
					  \mathcal{P}^e\,\rho&=\sum_{n=0}^\infty\,\ket{n}\bra{n}\,\langle n|\rho | n\rangle
 \end{split} \end{equation} 
with the external/vibrational density matrix contribution  $\mu=\mathcal{P}^e\,\text{tr}_I(\rho)$, the internal steady state $\rho_{ss}=\ket{D}\bra{D}$ and $\mathcal{P}^i$ and $\mathcal{P}^e$ the projectors on the internal steady state (assuming that the internal dynamics is much faster than the cooling dynamics) and the vibrational $\ket{n}\bra{n}$ subspace respectively. $\text{tr}_I$ denotes the trace over the internal part.
The master equation for the relevant external part $\mathcal{P}\rho$ up to first order in the Lamb-Dicke parameter is given by (see e.g. \cite{cirac95})
\begin{equation}\label{app7}  \dfrac{\mathrm{d}}{\mathrm{d}t}\mathcal{P}\rho= \mathcal{P}\mathcal{L}_1\,\left( -\mathcal{L}_0^{-1}\right)\mathcal{L}_1 \mathcal{P} \,\rho 
					=\int_{0}^\infty\,\mathrm{d}t \mathcal{P}\,\mathcal{L}_1\,e^{\mathcal{L}_0\,t}\,\mathcal{L}_1\mathcal{P}\rho\,.			  \end{equation}
Tracing equation (\ref{app7}) over the internal part, the master equation for the external part $\mu$ is obtained
\begin{equation}\label{app8} \dfrac{\mathrm{d}}{\mathrm{d}t}\,\mu  = \int_0^\infty \mathrm{d}t\,\,\mathcal{P}^e\,\text{tr}_I\,\left(\mathcal{L}_1\,e^{\mathcal{L}_0\,t}\mathcal{L}_1 \mathcal{P}\,\rho \right)
					=\int_0^\infty \mathrm{d}t\,\,\mathcal{P}^e\,\text{tr}_I\,\left(\mathcal{L}_1^1\,e^{\mathcal{L}_0\,t}\mathcal{L}_1^1 \mathcal{P}\,\rho \right) 			 \end{equation}
wherein it was used that $\text{tr}_I(\mathcal{P}\rho)=\mathcal{P}^e\,\text{tr}_I(\rho)$, which follows directly from eq.\,(\ref{app6}). The last step takes into account that the $\mathcal{L}_1^2$ terms (originating from the Schrieffer-Wolff transformation of the spontaneous decay terms) do not contribute in the master equation, because
\begin{equation}\label{app9} \mathcal{L}_1^2\,\mu\,\rho_{ss}=0,\qquad \mathcal{L}_1^2\rho_{ext}\rho_{int}=0   \end{equation}
with $\rho_{ss}=\ket{D}\bra{D}$ and furthermore assuming equal $\gamma$-rates in $\mathcal{L}_1^2$.\\
Inserting (\ref{app5}) into eq. (\ref{app3}) the first order Liouvillian $\mathcal{L}_1^1$ can be written as
\begin{equation}\label{app10}  \mathcal{L}_1^1\,\rho=-i\,\sum_n\,[F_n\,R_n,\rho]=-i\,\sum_n F_n\,\mathcal{L}_{1,ext}^n\,\rho+\mathcal{L}_{1,int}^n\,\rho\,R_n
 \end{equation}
using the definitions $\mathcal{L}_{1,ext}^n\,\rho=[R_n,\rho]$ and $\mathcal{L}_{1,int}^n\,\rho=[F_n,\rho]$.
Thus, inserting (\ref{app10}) into (\ref{app8}) and noting that
\begin{equation}\label{app11}  \text{tr}_I(\mathcal{L}_{1,int}^n\,\rho_{ext}\otimes\rho_{int})=\rho_{ext}\,\text{tr}_I([F_n,\rho_{int}])=0   \end{equation}
one obtains
\begin{equation}\label{app12}
\begin{split}
 \dfrac{\mathrm{d}}{\mathrm{d}t}\mu&= (-i)\,\sum_n \int_0^\infty \mathrm{d}t\,\mathcal{P}^e\text{tr}_I\left( F_n\,\mathcal{L}_{1,ext}^n\,e^{\mathcal{L}_0\,t}\,\mathcal{L}_1^1\,\mu\,\rho_{ss}  \right) \\
			&=(-1)\sum_{n,n'} \int_0^\infty \mathrm{d}t\mathcal{P}^e \mathcal{L}_{1,ext}^n e^{\mathcal{L}_{0,ext} t} \Bigl\{\mathcal{L}_{1,ext}^{n'}\, \mu\, \text{tr}_I\left(F_n\, e^{\mathcal{L}_{0,int} t}\, F_{n'}\, \rho_{ss}   \right)+\mu\,R_{n'}\,\text{tr}_I\left( F_n\,e^{\mathcal{L}_{0,int}\,t}\mathcal{L}_{1,int}^{n'}\,\rho_{ss} \right)  \Bigr\}\\
			&=(-1)\,\sum_{n,n'} \int_0^\infty \mathrm{d}t\,\mathcal{P}^e \mathcal{L}_{1,ext}^n\,e^{\mathcal{L}_{0,ext}\,t}\,\Bigl\{ \mathcal{L}_{1,ext}^{n'}\,\mu\,\langle F_{n}(t)\,F_{n'}(0) \rangle_{ss}+\mu\,R_{n'}\,\langle \left[ F_{n}(t),F_{n'}(0)  \right]  \rangle_{ss}  \Bigr\}\,,
\end{split}
   \end{equation}
with the internal operator two time-correlation functions $\langle F_n(t)\,F_{n'}(0)\rangle_{ss}=\text{tr}_I\left(F_n\, e^{\mathcal{L}_{0,int} t}\, F_{n'}\, \rho_{ss}   \right)$ and $\langle F_{n'}(0)\,F_n(t)\rangle_{ss}=\text{tr}_I\left(F_n\, e^{\mathcal{L}_{0,int} t}\, \rho_{ss} \,  F_{n'} \right)$.\\
Furthermore, using that $\left[\text{exp}(\mathcal{L}_{o,ext}\,t),\mu\right]=0$ and defining $R_n(t)=e^{\mathcal{L}_{0,ext}\,t}\,R_n$ we end up with
\begin{equation}\label{app13}  \dfrac{\mathrm{d}}{\mathrm{d}t}\,\mu=(-1)\sum_{n,n'} \int_0^\infty \mathrm{d}t\,\mathcal{P}^e\, \left\{ \left[ R_{n},[R_{n'}(t),\mu] \right]\,\langle F_{n}(t)\,F_{n'}(0)  \rangle_{ss}+\left[ R_{n},\mu\,R_{n'}(t) \right]\, \left[ F_{n}(t),F_{n'}(0)  \right]   \right\}\,.   \end{equation}
Inserting the concrete expressions for the external and internal operators (see definitions below eq.\,(\ref{app5})) and applying the projection $\mathcal{P}^e$ (eq.\,(\ref{app6})) the master equation for the external part turns out to be
\begin{equation}\label{app14} \dfrac{\mathrm{d}}{\mathrm{d}t}\,\mu=  S_{12}(-\nu)\,\left(b^\dagger\,\mu\,b-b\,b^\dagger\,\mu \right)+S_{21}(\nu)\,\left( b\,\mu\,b^\dagger-b^\dagger\,b\,\mu \right) +\text{h.c.} \end{equation}
wherein 
\begin{equation}\label{app15} S_{ij}(\nu)=\int_{0}^\infty \mathrm{d}t \,e^{i\nu t}\,\langle F_i(t)\,F_j(0)  \rangle_{ss}   \,. \end{equation}
The time evolution of the vibrational state population ($\mu=\sum_{n=0}^\infty p_n\,\ket{n}\bra{n}$) following from eq.\,(\ref{app14}) takes the form
\begin{equation}\label{app16} \dfrac{\mathrm{d}}{\mathrm{d}t}\,p_n=\bra{n}\dfrac{\mathrm{d}}{\mathrm{d}t}\mu\ket{n} 
  =(n+1)\,A_{-}\,p_{n+1}-\left[(n+1)\,A_{+}+n\,A_{-}  \right]\,p_n+n\,A_{+}\,p_{n-1}  \end{equation}
with the phonon number increasing / decreasing rates
\begin{equation}\label{app17}  A_-=2\,\text{Re}(S_{21}(\nu)), \qquad  A_+=2\,\text{Re}(S_{12}(-\nu)) \, .    \end{equation}
Therefore the equation of motion for the phonon population expectation value $\langle n \rangle$ is found to be
\begin{equation}\label{app18}   \dfrac{\mathrm{d}}{\mathrm{d}t}\langle n \rangle =\dfrac{\mathrm{d}}{\mathrm{d}t}\sum_{n=0}^\infty p_n\,n=\sum_{n=0}^\infty n\,\dfrac{\mathrm{d}}{\mathrm{d}t}p_n  
	=-(A_--A_+)\,\langle n \rangle+A_+  \end{equation}
with the steady state solution $\langle n\rangle_{ss}$
\begin{equation}\label{app19} \langle n \rangle_{ss}=\dfrac{A_+}{A_--A_+}\,.   \end{equation}
Thus it remains calculating the rates $A_-$ and $A_+$, i.e. calculating the two-time correlation functions $S_{ij}(\nu)$ (\ref{app15}). Using the Bloch equations (the internal equation of motion in the zeroth order in the Lamb-Dicke parameter) and the quantum regression theorem, one obtains the results in section \ref{sec_formula}.

\bibliography{lasercool}

\providecommand{\noopsort}[1]{}\providecommand{\singleletter}[1]{#1}%
\begin{thebibliography}{21}%
\makeatletter
\providecommand \@ifxundefined [1]{%
 \@ifx{#1\undefined}
}%
\providecommand \@ifnum [1]{%
 \ifnum #1\expandafter \@firstoftwo
 \else \expandafter \@secondoftwo
 \fi
}%
\providecommand \@ifx [1]{%
 \ifx #1\expandafter \@firstoftwo
 \else \expandafter \@secondoftwo
 \fi
}%
\providecommand \natexlab [1]{#1}%
\providecommand \enquote  [1]{``#1''}%
\providecommand \bibnamefont  [1]{#1}%
\providecommand \bibfnamefont [1]{#1}%
\providecommand \citenamefont [1]{#1}%
\providecommand \href@noop [0]{\@secondoftwo}%
\providecommand \href [0]{\begingroup \@sanitize@url \@href}%
\providecommand \@href[1]{\@@startlink{#1}\@@href}%
\providecommand \@@href[1]{\endgroup#1\@@endlink}%
\providecommand \@sanitize@url [0]{\catcode `\\12\catcode `\$12\catcode
  `\&12\catcode `\#12\catcode `\^12\catcode `\_12\catcode `\%12\relax}%
\providecommand \@@startlink[1]{}%
\providecommand \@@endlink[0]{}%
\providecommand \url  [0]{\begingroup\@sanitize@url \@url }%
\providecommand \@url [1]{\endgroup\@href {#1}{\urlprefix }}%
\providecommand \urlprefix  [0]{URL }%
\providecommand \Eprint [0]{\href }%
\@ifxundefined \urlstyle {%
  \providecommand \doi  [0]{\begingroup \@sanitize@url \@doi}%
  \providecommand \@doi [1]{\endgroup \@@startlink {\doibase
  #1}doi:\discretionary {}{}{}#1\@@endlink }%
}{%
  \providecommand \doi  [0]{doi:\discretionary{}{}{}\begingroup
  \urlstyle{rm}\Url }%
}%
\providecommand \doibase [0]{http://dx.doi.org/}%
\providecommand \Doi [0]{\begingroup \@sanitize@url \@Doi }%
\providecommand \@Doi  [1]{\endgroup\@@startlink{\doibase#1}\@@Doi}%
\providecommand \@@Doi [1]{#1\@@endlink}%
\providecommand \selectlanguage [0]{\@gobble}%
\providecommand \bibinfo  [0]{\@secondoftwo}%
\providecommand \bibfield  [0]{\@secondoftwo}%
\providecommand \translation [1]{[#1]}%
\providecommand \BibitemOpen [0]{}%
\providecommand \bibitemStop [0]{}%
\providecommand \bibitemNoStop [0]{.\EOS\space}%
\providecommand \EOS [0]{\spacefactor3000\relax}%
\providecommand \BibitemShut  [1]{\csname bibitem#1\endcsname}%
\bibitem [{\citenamefont {Wineland}\ \emph {et~al.}(2005)\citenamefont
  {Wineland}, \citenamefont {Leibfried}, \citenamefont {Barrett}, \citenamefont
  {Ben-Kish}, \citenamefont {Bergquist}, \citenamefont {Blakestad},
  \citenamefont {Bollinger}, \citenamefont {Britton}, \citenamefont
  {Chiaverini}, \citenamefont {Demarco}, \citenamefont {Hume}, \citenamefont
  {Itano}, \citenamefont {Jensen}, \citenamefont {Jost}, \citenamefont {Knill},
  \citenamefont {Koelemeij}, \citenamefont {Langer}, \citenamefont {Oskay},
  \citenamefont {Ozeri}, \citenamefont {Reichle}, \citenamefont {Rosenband},
  \citenamefont {Schaetz}, \citenamefont {Schmidt},\ and\ \citenamefont
  {Seidelin}}]{Wineland2005}%
  \BibitemOpen
  \bibfield  {author} {\bibinfo {author} {\bibfnamefont {D.~J.}\ \bibnamefont
  {Wineland}}, \bibinfo {author} {\bibfnamefont {D.}~\bibnamefont {Leibfried}},
  \bibinfo {author} {\bibfnamefont {M.~D.}\ \bibnamefont {Barrett}}, \bibinfo
  {author} {\bibfnamefont {A.}~\bibnamefont {Ben-Kish}}, \bibinfo {author}
  {\bibfnamefont {J.~C.}\ \bibnamefont {Bergquist}}, \bibinfo {author}
  {\bibfnamefont {R.~B.}\ \bibnamefont {Blakestad}}, \bibinfo {author}
  {\bibfnamefont {J.~J.}\ \bibnamefont {Bollinger}}, \bibinfo {author}
  {\bibfnamefont {J.}~\bibnamefont {Britton}}, \bibinfo {author} {\bibfnamefont
  {J.}~\bibnamefont {Chiaverini}}, \bibinfo {author} {\bibfnamefont
  {B.}~\bibnamefont {Demarco}}, \bibinfo {author} {\bibfnamefont
  {D.}~\bibnamefont {Hume}}, \bibinfo {author} {\bibfnamefont {W.~M.}\
  \bibnamefont {Itano}}, \bibinfo {author} {\bibfnamefont {M.}~\bibnamefont
  {Jensen}}, \bibinfo {author} {\bibfnamefont {J.~D.}\ \bibnamefont {Jost}},
  \bibinfo {author} {\bibfnamefont {E.}~\bibnamefont {Knill}}, \bibinfo
  {author} {\bibfnamefont {J.}~\bibnamefont {Koelemeij}}, \bibinfo {author}
  {\bibfnamefont {C.}~\bibnamefont {Langer}}, \bibinfo {author} {\bibfnamefont
  {W.}~\bibnamefont {Oskay}}, \bibinfo {author} {\bibfnamefont
  {R.}~\bibnamefont {Ozeri}}, \bibinfo {author} {\bibfnamefont
  {R.}~\bibnamefont {Reichle}}, \bibinfo {author} {\bibfnamefont
  {T.}~\bibnamefont {Rosenband}}, \bibinfo {author} {\bibfnamefont
  {T.}~\bibnamefont {Schaetz}}, \bibinfo {author} {\bibfnamefont {P.~O.}\
  \bibnamefont {Schmidt}}, \ and\ \bibinfo {author} {\bibfnamefont
  {S.}~\bibnamefont {Seidelin}},\ }\href@noop {} {\bibfield  {journal}
  {\bibinfo  {journal} {Proceedings of the International Conference on Laser
  Spectroscopy (ICOLS)}} (\bibinfo {year} {2005})}\BibitemShut {NoStop}%
\bibitem [{\citenamefont {Sasura}\ and\ \citenamefont
  {Buzek}(2002)}]{sasura01}%
  \BibitemOpen
  \bibfield  {author} {\bibinfo {author} {\bibfnamefont {M.}~\bibnamefont
  {Sasura}}\ and\ \bibinfo {author} {\bibfnamefont {V.}~\bibnamefont {Buzek}},\
  }\href@noop {} {\bibfield  {journal} {\bibinfo  {journal} {J. Mod. Opt.},\
  }\textbf {\bibinfo {volume} {49}},\ \bibinfo {pages} {1593} (\bibinfo {year}
  {2002})}\BibitemShut {NoStop}%
\bibitem [{\citenamefont {H{\"a}nsch}\ and\ \citenamefont
  {Schawlow}(1975)}]{haensch74}%
  \BibitemOpen
  \bibfield  {author} {\bibinfo {author} {\bibfnamefont {T.~W.}\ \bibnamefont
  {H{\"a}nsch}}\ and\ \bibinfo {author} {\bibfnamefont {A.~L.}\ \bibnamefont
  {Schawlow}},\ }\href@noop {} {\bibfield  {journal} {\bibinfo  {journal} {Opt.
  Commun.},\ }\textbf {\bibinfo {volume} {13}},\ \bibinfo {pages} {68}
  (\bibinfo {year} {1975})}\BibitemShut {NoStop}%
\bibitem [{\citenamefont {Wineland}\ and\ \citenamefont
  {Dehmelt}(1975)}]{wineland75}%
  \BibitemOpen
  \bibfield  {author} {\bibinfo {author} {\bibfnamefont {D.~J.}\ \bibnamefont
  {Wineland}}\ and\ \bibinfo {author} {\bibfnamefont {H.}~\bibnamefont
  {Dehmelt}},\ }\href@noop {} {\bibfield  {journal} {\bibinfo  {journal} {Bull.
  Am. Phys. Soc.},\ }\textbf {\bibinfo {volume} {20}},\ \bibinfo {pages} {637}
  (\bibinfo {year} {1975})}\BibitemShut {NoStop}%
\bibitem [{\citenamefont {Stenholm}(1986)}]{stenholm86}%
  \BibitemOpen
  \bibfield  {author} {\bibinfo {author} {\bibfnamefont {S.}~\bibnamefont
  {Stenholm}},\ }\href@noop {} {\bibfield  {journal} {\bibinfo  {journal} {Rev.
  Mod. Phys.},\ }\textbf {\bibinfo {volume} {58}},\ \bibinfo {pages} {699}
  (\bibinfo {year} {1986})}\BibitemShut {NoStop}%
\bibitem [{\citenamefont {Wineland}\ \emph {et~al.}(1978)\citenamefont
  {Wineland}, \citenamefont {Drullinger},\ and\ \citenamefont
  {Walls}}]{wineland78}%
  \BibitemOpen
  \bibfield  {author} {\bibinfo {author} {\bibfnamefont {D.~J.}\ \bibnamefont
  {Wineland}}, \bibinfo {author} {\bibfnamefont {R.~E.}\ \bibnamefont
  {Drullinger}}, \ and\ \bibinfo {author} {\bibfnamefont {F.~L.}\ \bibnamefont
  {Walls}},\ }\href@noop {} {\bibfield  {journal} {\bibinfo  {journal} {Phys.\
  Rev. Lett.},\ }\textbf {\bibinfo {volume} {40}},\ \bibinfo {pages} {1639}
  (\bibinfo {year} {1978})}\BibitemShut {NoStop}%
\bibitem [{\citenamefont {Lindberg}\ and\ \citenamefont
  {Javanainen}(1986)}]{lindberg86}%
  \BibitemOpen
  \bibfield  {author} {\bibinfo {author} {\bibfnamefont {M.}~\bibnamefont
  {Lindberg}}\ and\ \bibinfo {author} {\bibfnamefont {J.}~\bibnamefont
  {Javanainen}},\ }\href@noop {} {\bibfield  {journal} {\bibinfo  {journal}
  {J.Opt.Soc.Am.B},\ }\textbf {\bibinfo {volume} {3}},\ \bibinfo {pages} {1008}
  (\bibinfo {year} {1986})}\BibitemShut {NoStop}%
\bibitem [{\citenamefont {Kerman}\ \emph {et~al.}(2000)\citenamefont {Kerman},
  \citenamefont {Vuletic}, \citenamefont {Chin},\ and\ \citenamefont
  {Chu}}]{kerman00}%
  \BibitemOpen
  \bibfield  {author} {\bibinfo {author} {\bibfnamefont {A.~J.}\ \bibnamefont
  {Kerman}}, \bibinfo {author} {\bibfnamefont {V.}~\bibnamefont {Vuletic}},
  \bibinfo {author} {\bibfnamefont {C.}~\bibnamefont {Chin}}, \ and\ \bibinfo
  {author} {\bibfnamefont {S.}~\bibnamefont {Chu}},\ }\href@noop {} {\bibfield
  {journal} {\bibinfo  {journal} {Phys.\ Rev. Lett.},\ }\textbf {\bibinfo
  {volume} {84}},\ \bibinfo {pages} {439} (\bibinfo {year} {2000})}\BibitemShut
  {NoStop}%
\bibitem [{\citenamefont {Lee}\ \emph {et~al.}(1996)\citenamefont {Lee},
  \citenamefont {Adams}, \citenamefont {Kasevich},\ and\ \citenamefont
  {Chu}}]{lee96}%
  \BibitemOpen
  \bibfield  {author} {\bibinfo {author} {\bibfnamefont {H.~J.}\ \bibnamefont
  {Lee}}, \bibinfo {author} {\bibfnamefont {C.~S.}\ \bibnamefont {Adams}},
  \bibinfo {author} {\bibfnamefont {M.}~\bibnamefont {Kasevich}}, \ and\
  \bibinfo {author} {\bibfnamefont {S.}~\bibnamefont {Chu}},\ }\href@noop {}
  {\bibfield  {journal} {\bibinfo  {journal} {Phys.\ Rev. Lett.},\ }\textbf
  {\bibinfo {volume} {76}},\ \bibinfo {pages} {2658} (\bibinfo {year}
  {1996})}\BibitemShut {NoStop}%
\bibitem [{\citenamefont {Monroe}\ \emph {et~al.}(1995)\citenamefont {Monroe},
  \citenamefont {Meekhof}, \citenamefont {King}, \citenamefont {Jefferts},
  \citenamefont {Itano},\ and\ \citenamefont {Wineland}}]{monroe95}%
  \BibitemOpen
  \bibfield  {author} {\bibinfo {author} {\bibfnamefont {C.}~\bibnamefont
  {Monroe}}, \bibinfo {author} {\bibfnamefont {D.~M.}\ \bibnamefont {Meekhof}},
  \bibinfo {author} {\bibfnamefont {B.~E.}\ \bibnamefont {King}}, \bibinfo
  {author} {\bibfnamefont {S.~R.}\ \bibnamefont {Jefferts}}, \bibinfo {author}
  {\bibfnamefont {W.~M.}\ \bibnamefont {Itano}}, \ and\ \bibinfo {author}
  {\bibfnamefont {D.~J.}\ \bibnamefont {Wineland}},\ }\href@noop {} {\bibfield
  {journal} {\bibinfo  {journal} {Phys.\ Rev. Lett.},\ }\textbf {\bibinfo
  {volume} {75}},\ \bibinfo {pages} {4011} (\bibinfo {year}
  {1995})}\BibitemShut {NoStop}%
\bibitem [{\citenamefont {Morigi}\ \emph {et~al.}(2000)\citenamefont {Morigi},
  \citenamefont {Eschner},\ and\ \citenamefont {Keitel}}]{morigi00}%
  \BibitemOpen
  \bibfield  {author} {\bibinfo {author} {\bibfnamefont {G.}~\bibnamefont
  {Morigi}}, \bibinfo {author} {\bibfnamefont {J.}~\bibnamefont {Eschner}}, \
  and\ \bibinfo {author} {\bibfnamefont {C.~H.}\ \bibnamefont {Keitel}},\
  }\href@noop {} {\bibfield  {journal} {\bibinfo  {journal} {Phys.\ Rev.
  Lett.},\ }\textbf {\bibinfo {volume} {85}},\ \bibinfo {pages} {4458}
  (\bibinfo {year} {2000})}\BibitemShut {NoStop}%
\bibitem [{\citenamefont {Roos}\ \emph {et~al.}(2000)\citenamefont {Roos},
  \citenamefont {Leibfried}, \citenamefont {Mundt}, \citenamefont
  {Schmidt-Kaler}, \citenamefont {Eschner},\ and\ \citenamefont
  {Blatt}}]{roos00}%
  \BibitemOpen
  \bibfield  {author} {\bibinfo {author} {\bibfnamefont {C.~F.}\ \bibnamefont
  {Roos}}, \bibinfo {author} {\bibfnamefont {D.}~\bibnamefont {Leibfried}},
  \bibinfo {author} {\bibfnamefont {A.}~\bibnamefont {Mundt}}, \bibinfo
  {author} {\bibfnamefont {F.}~\bibnamefont {Schmidt-Kaler}}, \bibinfo {author}
  {\bibfnamefont {J.}~\bibnamefont {Eschner}}, \ and\ \bibinfo {author}
  {\bibfnamefont {R.}~\bibnamefont {Blatt}},\ }\href@noop {} {\bibfield
  {journal} {\bibinfo  {journal} {Phys.\ Rev. Lett.},\ }\textbf {\bibinfo
  {volume} {85}},\ \bibinfo {pages} {5547} (\bibinfo {year}
  {2000})}\BibitemShut {NoStop}%
\bibitem [{\citenamefont {Fleischhauer}\ \emph {et~al.}(2005)\citenamefont
  {Fleischhauer}, \citenamefont {Imamoglu},\ and\ \citenamefont
  {Marangos}}]{fleischhauer05}%
  \BibitemOpen
  \bibfield  {author} {\bibinfo {author} {\bibfnamefont {M.}~\bibnamefont
  {Fleischhauer}}, \bibinfo {author} {\bibfnamefont {A.}~\bibnamefont
  {Imamoglu}}, \ and\ \bibinfo {author} {\bibfnamefont {J.~P.}\ \bibnamefont
  {Marangos}},\ }\href@noop {} {\bibfield  {journal} {\bibinfo  {journal} {Rev.
  Mod. Phys.},\ }\textbf {\bibinfo {volume} {77}},\ \bibinfo {pages} {633}
  (\bibinfo {year} {2005})}\BibitemShut {NoStop}%
\bibitem [{\citenamefont {Jonathan}\ \emph {et~al.}(2000)\citenamefont
  {Jonathan}, \citenamefont {Plenio},\ and\ \citenamefont
  {Knight}}]{jonathan00}%
  \BibitemOpen
  \bibfield  {author} {\bibinfo {author} {\bibfnamefont {D.}~\bibnamefont
  {Jonathan}}, \bibinfo {author} {\bibfnamefont {M.~B.}\ \bibnamefont
  {Plenio}}, \ and\ \bibinfo {author} {\bibfnamefont {P.~L.}\ \bibnamefont
  {Knight}},\ }\href@noop {} {\bibfield  {journal} {\bibinfo  {journal} {Phys.
  Rev. A.},\ }\textbf {\bibinfo {volume} {62}},\ \bibinfo {pages} {042307}
  (\bibinfo {year} {2000})}\BibitemShut {NoStop}%
\bibitem [{\citenamefont {Retzker}\ and\ \citenamefont
  {Plenio}(2007)}]{retzker07}%
  \BibitemOpen
  \bibfield  {author} {\bibinfo {author} {\bibfnamefont {A.}~\bibnamefont
  {Retzker}}\ and\ \bibinfo {author} {\bibfnamefont {M.~B.}\ \bibnamefont
  {Plenio}},\ }\href@noop {} {\bibfield  {journal} {\bibinfo  {journal} {New J.
  Phys.},\ }\textbf {\bibinfo {volume} {9}},\ \bibinfo {pages} {279} (\bibinfo
  {year} {2007})}\BibitemShut {NoStop}%
\bibitem [{\citenamefont {Cerrillo}\ \emph {et~al.}(2010)\citenamefont
  {Cerrillo}, \citenamefont {Retzker},\ and\ \citenamefont
  {Plenio}}]{cerrillo10}%
  \BibitemOpen
  \bibfield  {author} {\bibinfo {author} {\bibfnamefont {J.}~\bibnamefont
  {Cerrillo}}, \bibinfo {author} {\bibfnamefont {A.}~\bibnamefont {Retzker}}, \
  and\ \bibinfo {author} {\bibfnamefont {M.~B.}\ \bibnamefont {Plenio}},\
  }\href@noop {} {\bibfield  {journal} {\bibinfo  {journal} {Phys.\ Rev.
  Lett.},\ }\textbf {\bibinfo {volume} {104}},\ \bibinfo {pages} {043003}
  (\bibinfo {year} {2010})}\BibitemShut {NoStop}%
\bibitem [{\citenamefont {Mintert}\ and\ \citenamefont
  {Wunderlich}(2001)}]{mintert01}%
  \BibitemOpen
  \bibfield  {author} {\bibinfo {author} {\bibfnamefont {F.}~\bibnamefont
  {Mintert}}\ and\ \bibinfo {author} {\bibfnamefont {C.}~\bibnamefont
  {Wunderlich}},\ }\href@noop {} {\bibfield  {journal} {\bibinfo  {journal}
  {Phys.\ Rev. Lett.},\ }\textbf {\bibinfo {volume} {87}},\ \bibinfo {pages}
  {257904} (\bibinfo {year} {2001})}\BibitemShut {NoStop}%
\bibitem [{\citenamefont {Wunderlich}\ and\ \citenamefont
  {Balzer}(2003)}]{wunderlich03}%
  \BibitemOpen
  \bibfield  {author} {\bibinfo {author} {\bibfnamefont {C.}~\bibnamefont
  {Wunderlich}}\ and\ \bibinfo {author} {\bibfnamefont {C.}~\bibnamefont
  {Balzer}},\ }\href@noop {} {\bibfield  {journal} {\bibinfo  {journal} {Adv.
  Atom. Mol. Opt. Phys.},\ }\textbf {\bibinfo {volume} {49}},\ \bibinfo {pages}
  {293} (\bibinfo {year} {2003})}\BibitemShut {NoStop}%
\bibitem [{\citenamefont {Schrieffer}\ and\ \citenamefont
  {Wolff}(1966)}]{schrieffer66}%
  \BibitemOpen
  \bibfield  {author} {\bibinfo {author} {\bibfnamefont {J.~R.}\ \bibnamefont
  {Schrieffer}}\ and\ \bibinfo {author} {\bibfnamefont {P.~A.}\ \bibnamefont
  {Wolff}},\ }\href@noop {} {\bibfield  {journal} {\bibinfo  {journal} {Phys.
  Rev.},\ }\textbf {\bibinfo {volume} {149}},\ \bibinfo {pages} {491} (\bibinfo
  {year} {1966})}\BibitemShut {NoStop}%
\bibitem [{\citenamefont {Cirac}\ \emph {et~al.}(1992)\citenamefont {Cirac},
  \citenamefont {Blatt}, \citenamefont {Zoller},\ and\ \citenamefont
  {Phillips}}]{cirac92}%
  \BibitemOpen
  \bibfield  {author} {\bibinfo {author} {\bibfnamefont {J.~I.}\ \bibnamefont
  {Cirac}}, \bibinfo {author} {\bibfnamefont {R.}~\bibnamefont {Blatt}},
  \bibinfo {author} {\bibfnamefont {P.}~\bibnamefont {Zoller}}, \ and\ \bibinfo
  {author} {\bibfnamefont {W.~D.}\ \bibnamefont {Phillips}},\ }\href@noop {}
  {\bibfield  {journal} {\bibinfo  {journal} {Phys.\ Rev. A},\ }\textbf
  {\bibinfo {volume} {46}},\ \bibinfo {pages} {2668} (\bibinfo {year}
  {1992})}\BibitemShut {NoStop}%
\bibitem [{\citenamefont {Cirac}\ and\ \citenamefont {Zoller}(1995)}]{cirac95}%
  \BibitemOpen
  \bibfield  {author} {\bibinfo {author} {\bibfnamefont {J.~I.}\ \bibnamefont
  {Cirac}}\ and\ \bibinfo {author} {\bibfnamefont {P.}~\bibnamefont {Zoller}},\
  }\href@noop {} {\bibfield  {journal} {\bibinfo  {journal} {Phys.\ Rev.
  Lett.},\ }\textbf {\bibinfo {volume} {74}},\ \bibinfo {pages} {4091}
  (\bibinfo {year} {1995})}\BibitemShut {NoStop}%
\end{thebibliography}%

\end{document}